\documentclass[aps,
twocolumn,prl,
showpacs,preprintnumbers,nofootinbib,
amsmath,amssymb,
aps,
superscriptaddress]{revtex4-1}
\usepackage[breaklinks=true]{hyperref}
\usepackage{breakcites}
\usepackage{float}
\usepackage{graphicx}
\usepackage{dcolumn}
\usepackage{bm}
\usepackage{hyperref}
\usepackage{ulem} 
\usepackage{breakcites}

\newcommand{\be}{\begin{equation}}
\newcommand{\ee}{\end{equation}}
\newcommand{\bi}{\begin{itemize}}
\newcommand{\ei}{\end{itemize}}
\newcommand{\bea}{\begin{eqnarray}}
\newcommand{\eea}{\end{eqnarray}}

\usepackage{color}

\begin{document}

\title{Confusing head-on and precessing intermediate-mass binary black hole mergers}


\affiliation{Instituto Galego de F\'{i}sica de Altas Enerx\'{i}as, Universidade de Santiago de Compostela, 15782 Santiago de Compostela, Galicia, Spain}
\affiliation{Department of Physics, The Chinese University of Hong Kong, Shatin, N.T., Hong Kong}
\affiliation{Monash Centre for Astrophysics, School of Physics and Astronomy, Monash University, VIC 3800, Australia}
\affiliation{OzGrav: The ARC Centre of Excellence for Gravitational-Wave Discovery, Clayton, VIC 3800, Australia}
\affiliation{Centro de Astrof\'{i}sica e Gravita\c{c}\~{a}o - CENTRA,
Departamento de F\'{i}sica, Instituto Superior T\'{e}cnico - IST,Universidade de Lisboa - UL, Avenida Rovisco Pais 1, 1049-001, Portugal}
\affiliation{Departamento  de  Matem\'{a}tica  da  Universidade  de  Aveiro  and  Centre  for  Research  and  Development in  Mathematics  and  Applications  (CIDMA),  Campus  de  Santiago,  3810-183  Aveiro,  Portugal}
\affiliation{Max Planck Institute for Gravitational Physics (Albert Einstein Institute), Am M\"uhlenberg 1, Potsdam 14476, Germany}
\affiliation{Departamento de Astronom\'{i}a y Astrof\'{i}sica, Universitat de Val\`{e}ncia,
Dr. Moliner 50, 46100, Burjassot (Val\`{e}ncia), Spain}
\affiliation{Observatori Astron\`{o}mic, Universitat de Val\`{e}ncia,
C/ Catedr\'{a}tico Jos\'{e} Beltr\'{a}n 2, 46980, Paterna (Val\`{e}ncia), Spain}


\author{Juan Calder\'on~Bustillo$^{1,2,3,4}$}\noaffiliation
\author{Nicolas Sanchis-Gual$^{5,6}$}\noaffiliation
\author{Alejandro Torres-Forn\'e$^\text{7,8}$}\noaffiliation
\author{Jos\'e A. Font$^\text{8,9}$}\noaffiliation

\preprint{LIGO-P1900363}
\begin{abstract}


We report a degeneracy between the gravitational-wave signals from quasi-circular precessing black-hole mergers and those from extremely eccentric mergers, namely head-on collisions. Performing model selection on numerically simulated signals of head-on collisions using models for quasi-circular binaries we find that, for signal-to-noise ratios \textcolor{black}{of 15 and 25, typical of Advanced LIGO observations, head-on mergers with respective total masses of \textcolor{black}{$M\in (125,300)M_\odot$} and \textcolor{black}{$M\in (200,440)M_\odot$}} would be identified as precessing quasi-circular intermediate-mass black hole binaries, located at a much larger distance. Ruling out the head-on scenario would require to perform model selection using currently nonexistent waveform models for head-on collisions, together with the application of astrophysically motivated priors on the (rare) occurrence of those events. We show that in situations where standard parameter inference of compact binaries may report component masses inside (outside) the pair-instability supernova gap, the true object may be a head-on merger with masses outside (inside) this gap. We briefly discuss the potential implications of these findings for 
GW190521, \textcolor{black}{which we analyse in detail in \cite{Proca}.}

\end{abstract}
\maketitle


\paragraph*{\textbf{Introduction.--}}

To date, the Advanced LIGO-Virgo gravitational-wave (GW) detector network \cite{TheLIGOScientific:2014jea,TheVirgo:2014hva} has confirmed the observation of 
fifteen compact binary mergers \cite{Abbott:2016blz,Abbott:2016nmj,PhysRevLett.118.221101,Abbott:2017oio,Abbott:2017gyy,Abbott:2020uma,LIGOScientific:2020stg,TheLIGOScientific:2017qsa, BNS2, GW190521D,GW190521I}.
Some of these represent milestones for GW astronomy and science itself. The events GW150914 \cite{Abbott:2016blz} and GW170817 \cite{TheLIGOScientific:2017qsa} represented the first observation of a binary black hole (BBH) merger and a binary neutron star (BNS) merger, respectively, while the event GW190814 might be the first observation of a NS-BH merger \cite{NSBH}. Recently, LIGO-Virgo announced GW190521, a short signal consistent with the merger of two massive BHs in a quasi-circular orbit with signatures of orbital precession, at a distance of $d_L\simeq$ 5 Gpc, 
producing the first conclusively observed intermediate-mass BH (IMBH) with a mass of $\simeq142\,M_\odot$ \cite{GW190521D,GW190521I}. 

The most massive component of GW190521 convincingly populates the pair-instability supernova (PISN) mass gap ($\sim 65-130\,M_\odot$) predicted from stellar evolution 
of very massive, low-metallicity stars \cite{Heger:2002by}. Formation channels for BHs in the mass gap and above have been invoked in connection with PISN and pulsational PISN of progenitors with massive CO cores \cite{Woosley2002,Uchida2019} (see ~\cite{DiCarlo2019,Fragione1,Fragione2} for alternative proposals). Regardless of their formation, such collisions are the loudest sources for current detectors \cite{Abbott:2017iws} and, as exemplified by GW190521, lead to remnant BH masses in the lower end of the IMBH range $\sim100-250\,M_\odot$. 

\begin{table}[hb!]
\begin{tabular}{c|c|c|c|c|c|c|c}
$q=m_1/m_2$       & $a_1$ & $a_2$  & $D (M)$  & $e_{1}^{\rm red}$& $e_{2}^{\rm red}$ & $a_{\rm{fin}}$& $M_{\rm fin}/M_{\rm ini}$  \\ \hline
1& 0.58 &       0.58                   & 2.34 & 0.14 & 0.14 &0.35  &0.999 \\ 
2    & 0.60              &     0.56 & 2.34 &0.09 & 0.20   &0.39&0.999 \\ 
3     &    0.61          &   0.55 &2.34& 0.06  & 0.23 &0.43&0.999 \\ 
1     & 0.00            &    0.00    &  2.34&           0.00       & 0.00  & 0.00& 0.999 

\end{tabular}
\caption{Parameters of our head-on collision simulations: mass ratio ($q$), spins ($a_i$), separation ($D$), residual eccentricities ($e^{\rm red}_i$), final spin $a_{\rm{fin}}$, and mass loss ($M_{\rm fin}/M_{\rm ini}$).}
\label{models}
\end{table}

\begin{figure*}
\includegraphics[width=0.32\textwidth]{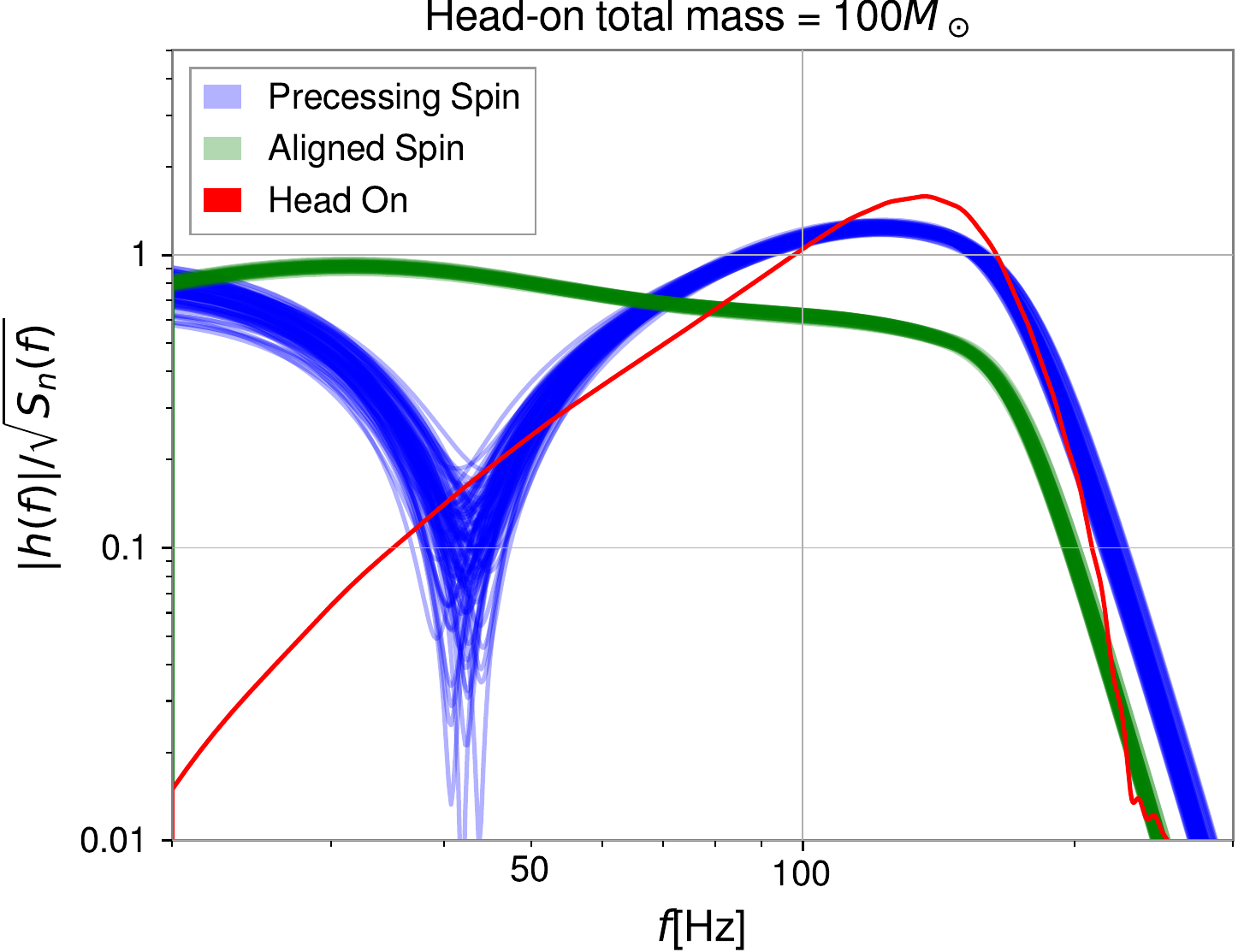}
\includegraphics[width=0.32\textwidth]{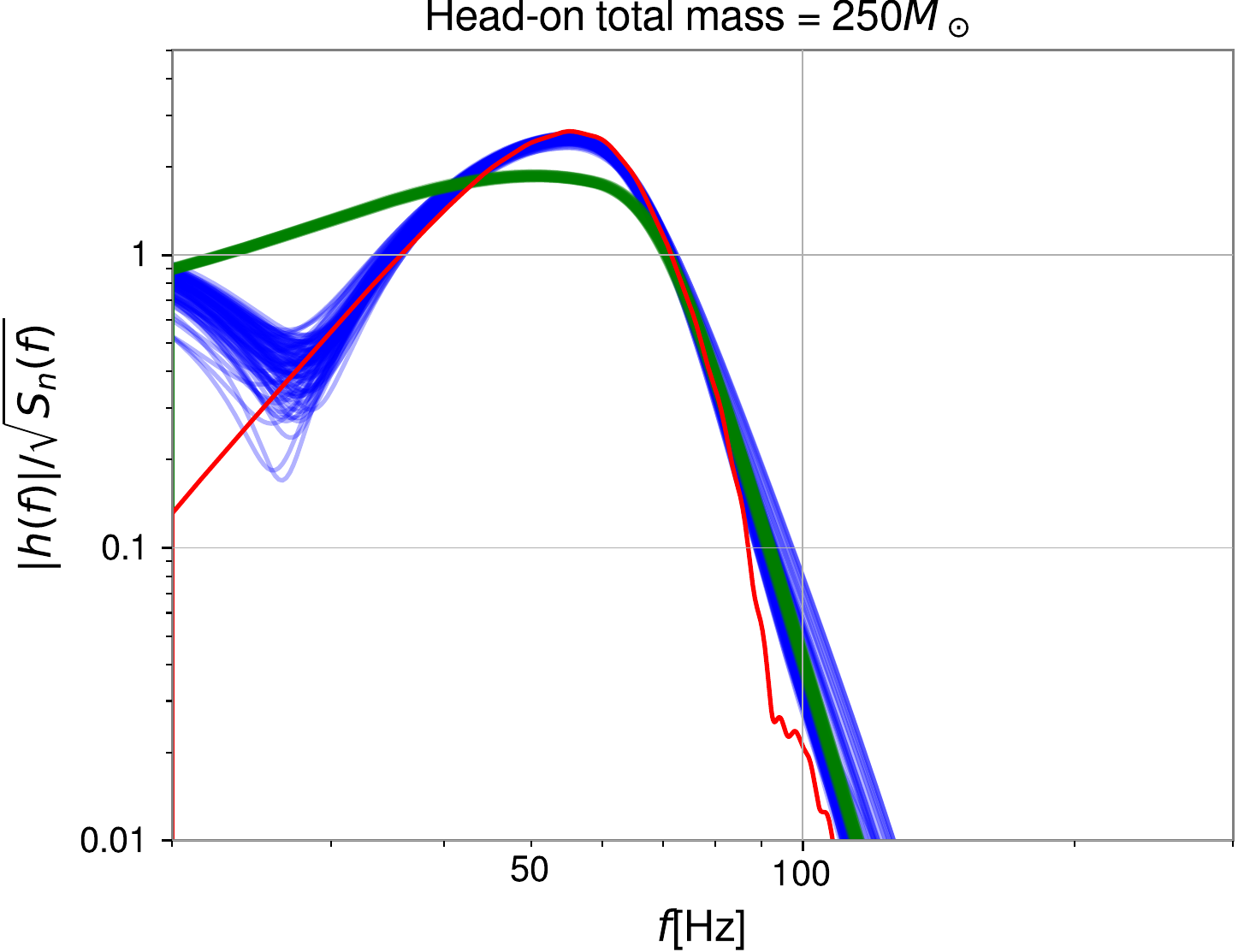}
\includegraphics[width=0.32\textwidth]{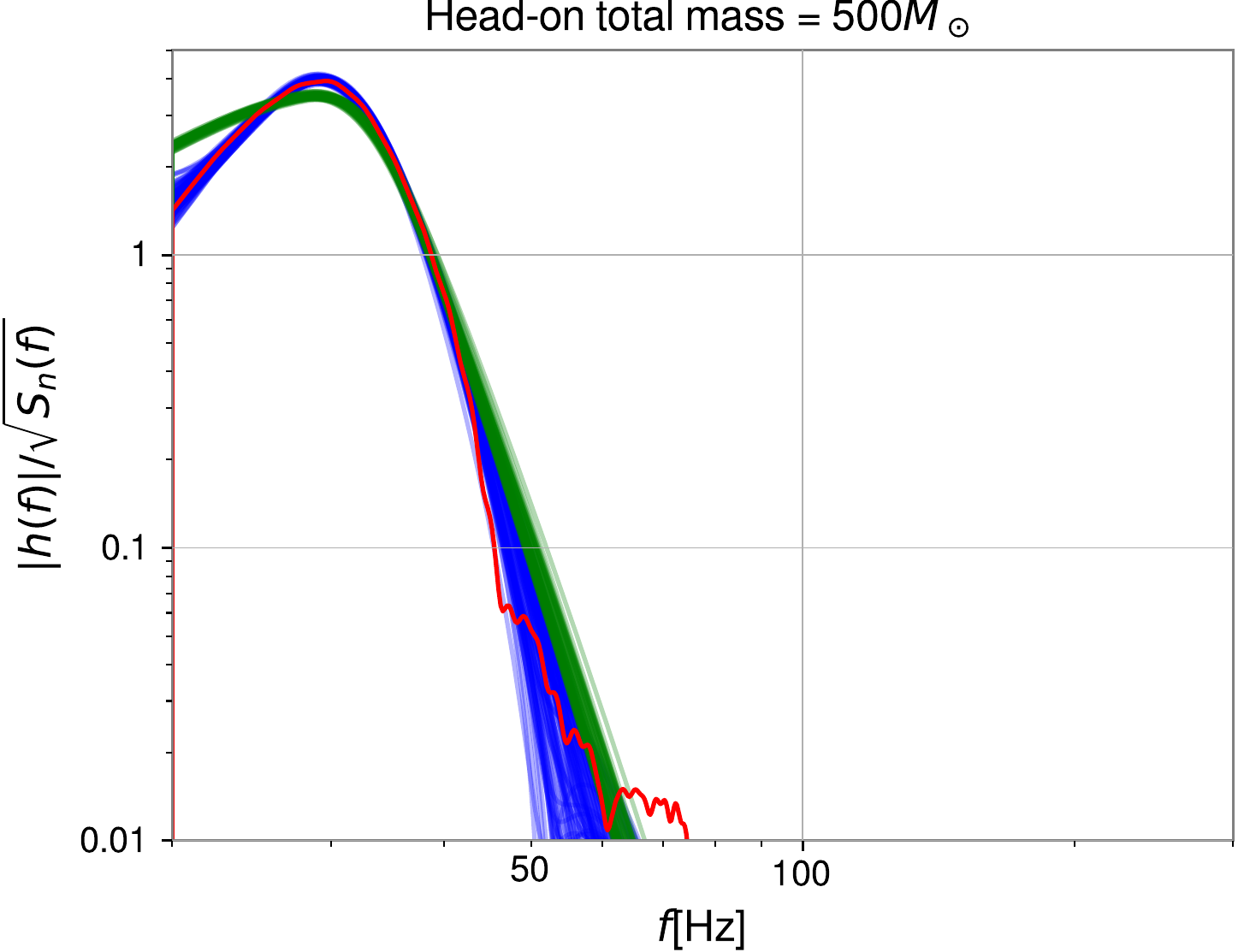}
\includegraphics[width=0.32\textwidth]{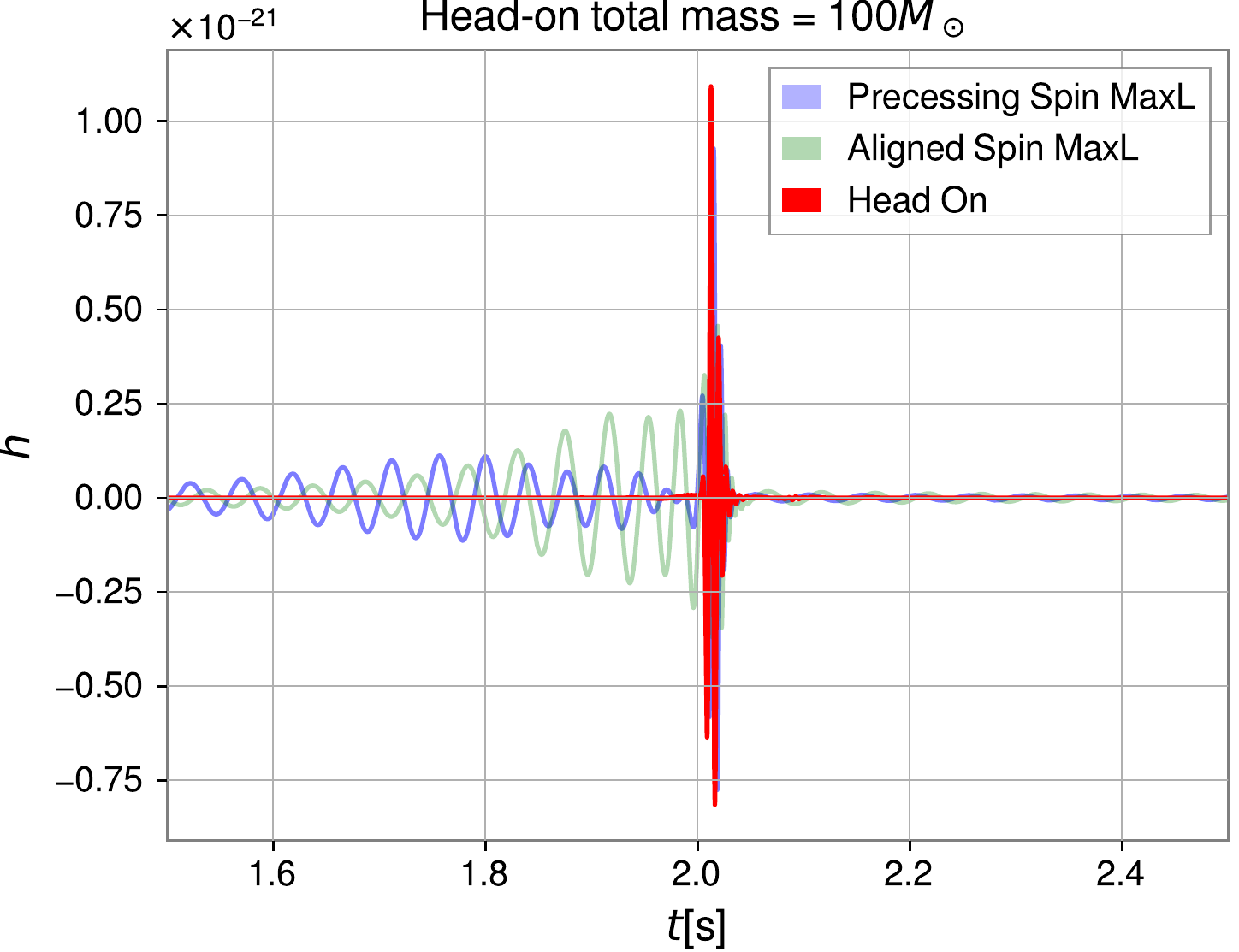}
\includegraphics[width=0.32\textwidth]{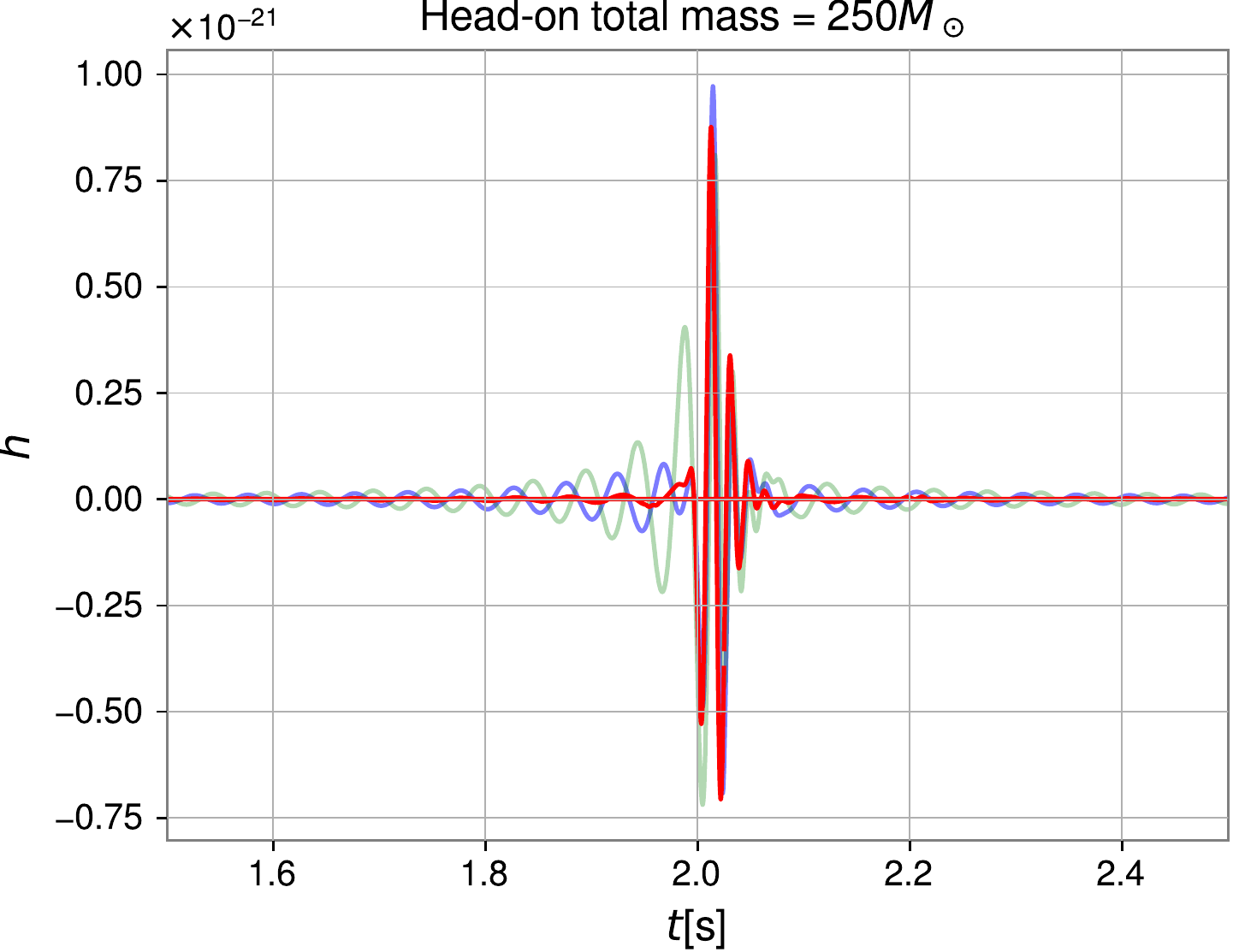}
\includegraphics[width=0.32\textwidth]{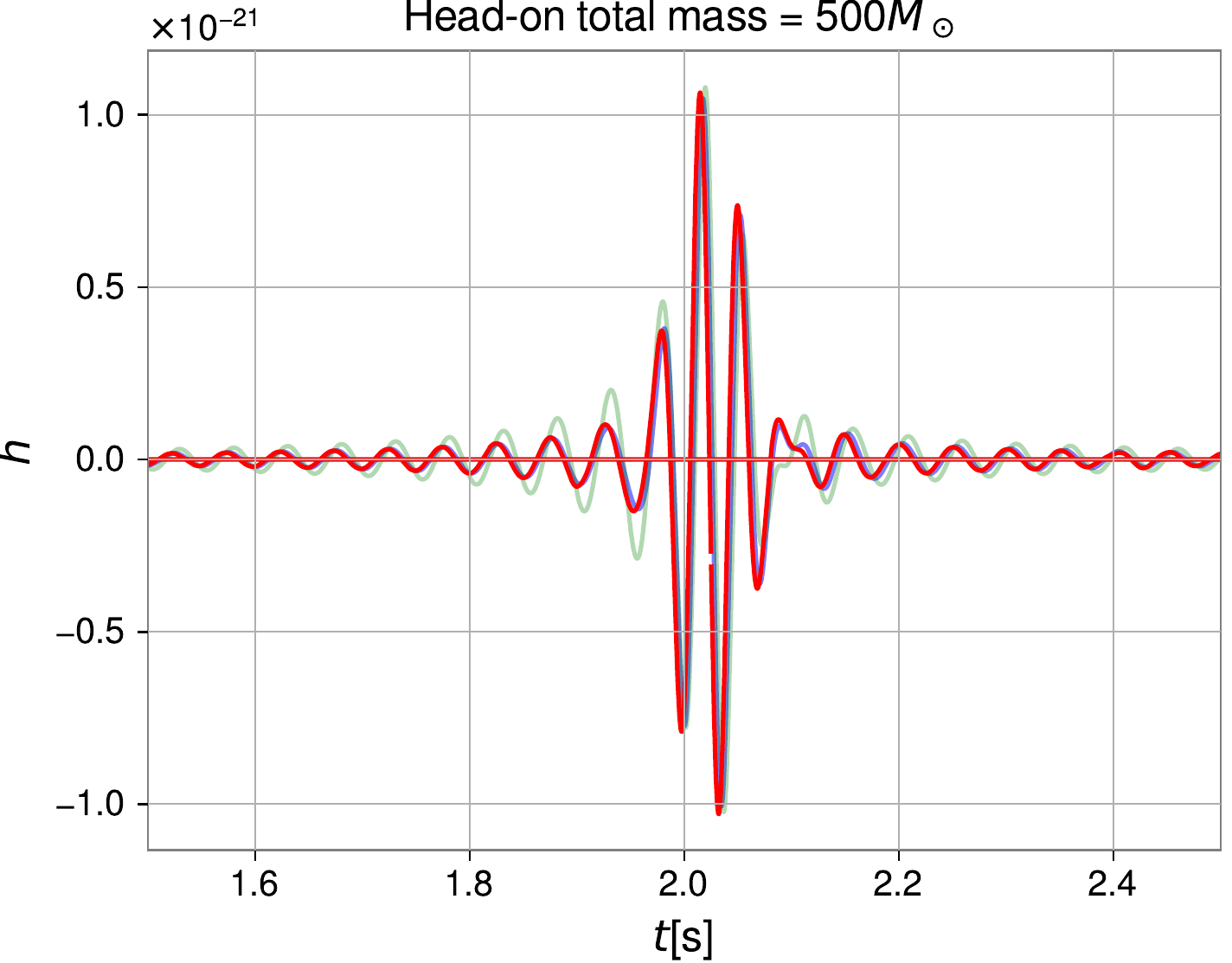}
\caption{Spectra and strain of HOCs and quasi-circular BBH mergers. Top: Amplitude of the Fourier transform of a $q=2$ HOC injection (red) with varying total mass, together with that of the best-matching posterior samples for aligned-spin (green) and precessing (blue) quasi-circular binaries. Bottom: Corresponding strains band-passed between the lower and upper frequency cutoffs of our analysis $f \in [20,512]$ Hz. 
}
\label{fig:spectra}
\end{figure*}

Due to the low frequency of such signals only the merger and ringdown portions yield significant power in the detector sensitive band, leaving a barely observable inspiral. If the latter were completely out of band, it might be impossible to determine the formation channel of the remnant, making it unsafe to assume a standard quasi-circular BBH origin. If, however, a few inspiral cycles are visible in band, it may still be possible to determine this origin. For instance, such cycles would be strong for a non-precessing BBH while the modulations induced by a strongly precessing orbital plane can significantly suppress the inspiral right before merger \cite{Healy2019,Boyle2019} causing a characteristic sine-Gaussian morphology. There are, however, alternative situations that may lead to such sine-Gaussian waveform, like hyperbolic encounters \cite{gold2013} and, as we will discuss, highly eccentric mergers \cite{Fragione3,Fragione4,Fragione5}.

Standard parameter estimation of LIGO-Virgo signals implements waveform templates for quasi-circular BBH mergers with negligible eccentricity. In this \textit{Letter} we investigate whether similarities in the waveform morphology may lead to a confusion between precessing and extremely eccentric BBH mergers (head-on collisions; HOCs), discussing the astrophysical implications. 
We estimate the mass range and signal loudness in which such confusion might occur. We show that for masses \textcolor{black}{$M\in (125,300)M_\odot$} and signal-to-noise ratios (SNRs) typical among current observations, precessing BBH waveforms can be confused with those from HOCs, producing a large over-estimation of the distance and a consequent under-estimation of the source-frame mass. Therefore, a HOC in this mass range could be genuinely identified as a precessing IMBH binary if only a quasi-circular merger scenario were considered. Remarkably, we find that if the HOC components are outside the PISN gap, parameter estimation with BBH waveforms may place one of the two masses inside it, reporting an apparent violation of the PISN gap.

\begin{figure}[hb!]
\includegraphics[width=1.0\columnwidth]{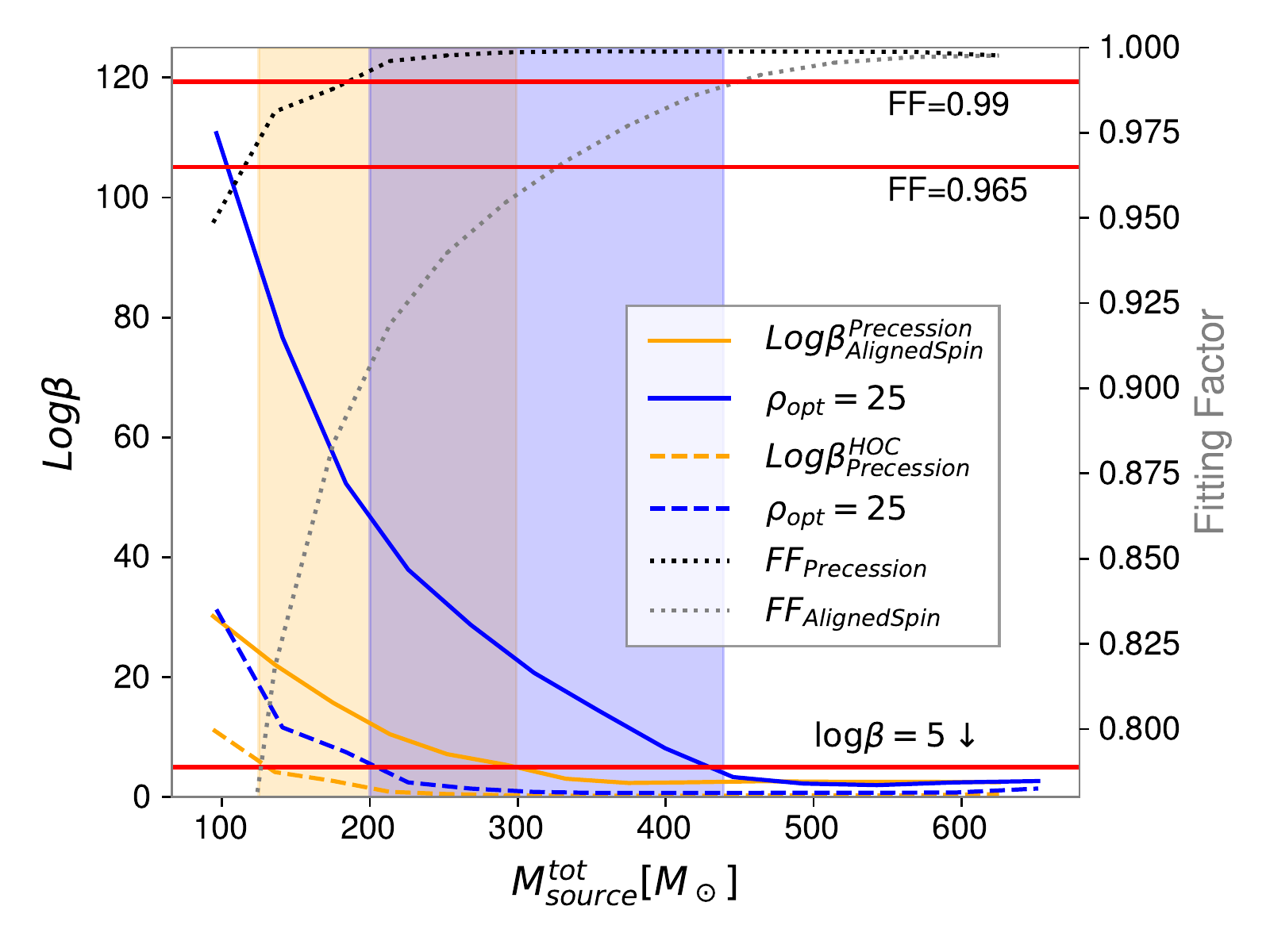}
\caption{LogEvidence $\log\beta^{{\tt BBH,IMRPhenomPv2}}_{{\tt BBH,IMRPhenomD}}$ (solid) and $\log\beta^{{\tt HOC}}_{{\tt BBH,IMRPhenomPv2}}$ (dashed) as a function of total source-frame mass and optimal SNR $\rho_{\rm opt}$ of the injected HOC signal (15-25 for orange-blue), for the $q=2$ case of Table I. The black and grey dotted lines denote the FF of the BBH models to our injection. The shaded regions denote the mass ranges where the HOC signal would be confused with a precessing BBH. Almost identical results are obtained for all of our simulations.
}
\label{fig:logB}
\end{figure}

\begin{figure}
\includegraphics[width=0.236\textwidth]{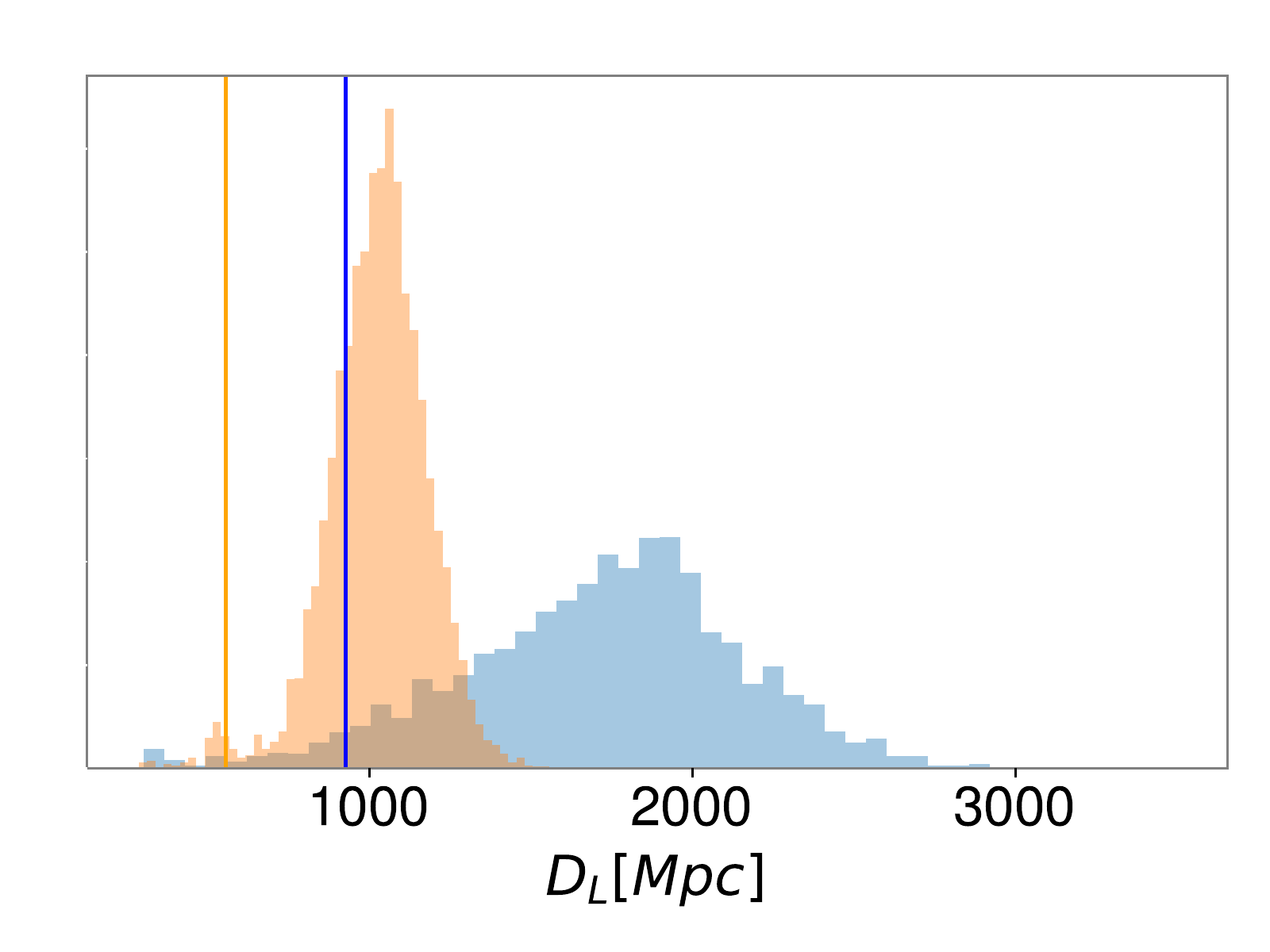}
\includegraphics[width=0.23\textwidth]{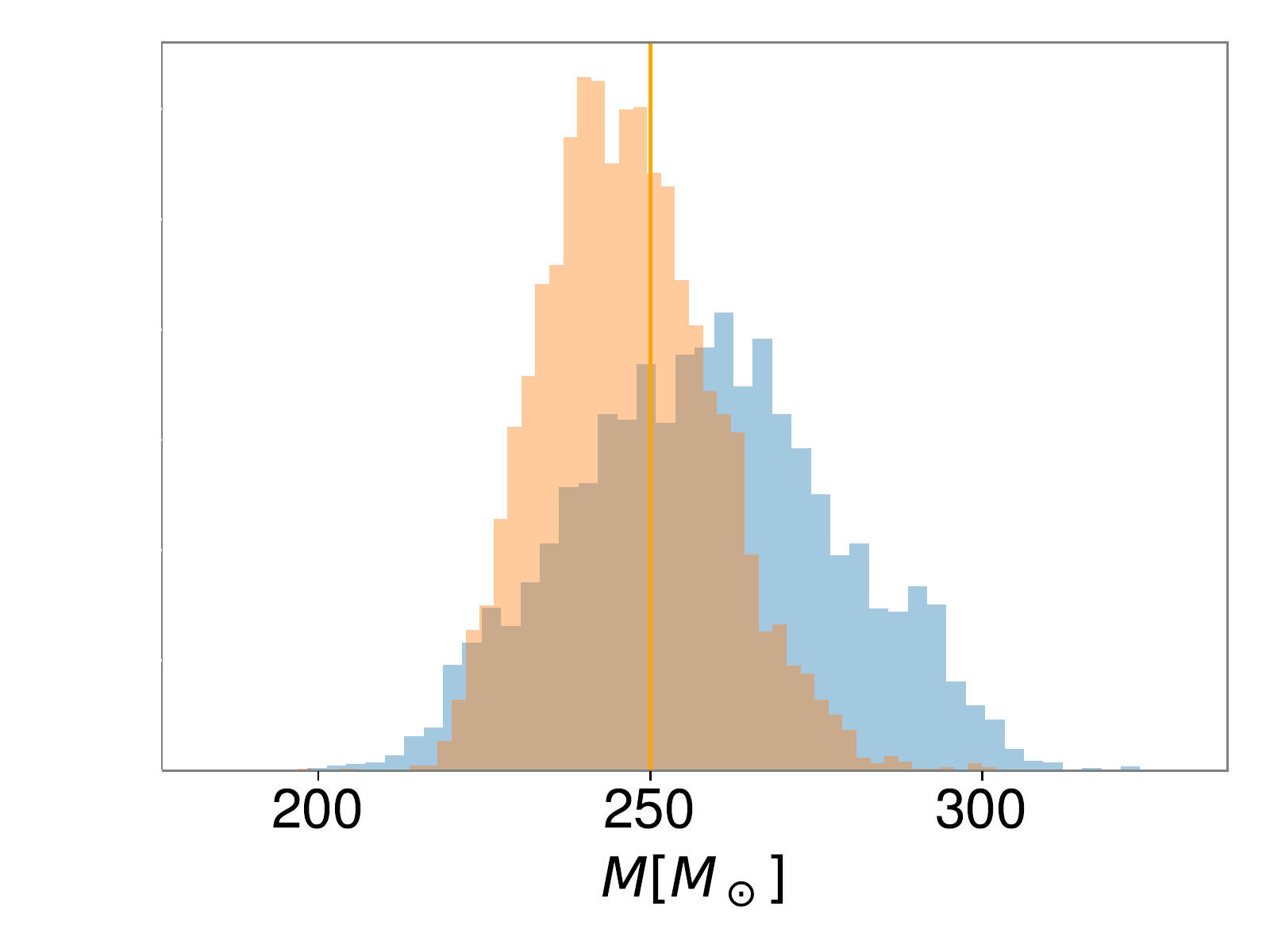}
\includegraphics[width=0.23\textwidth]{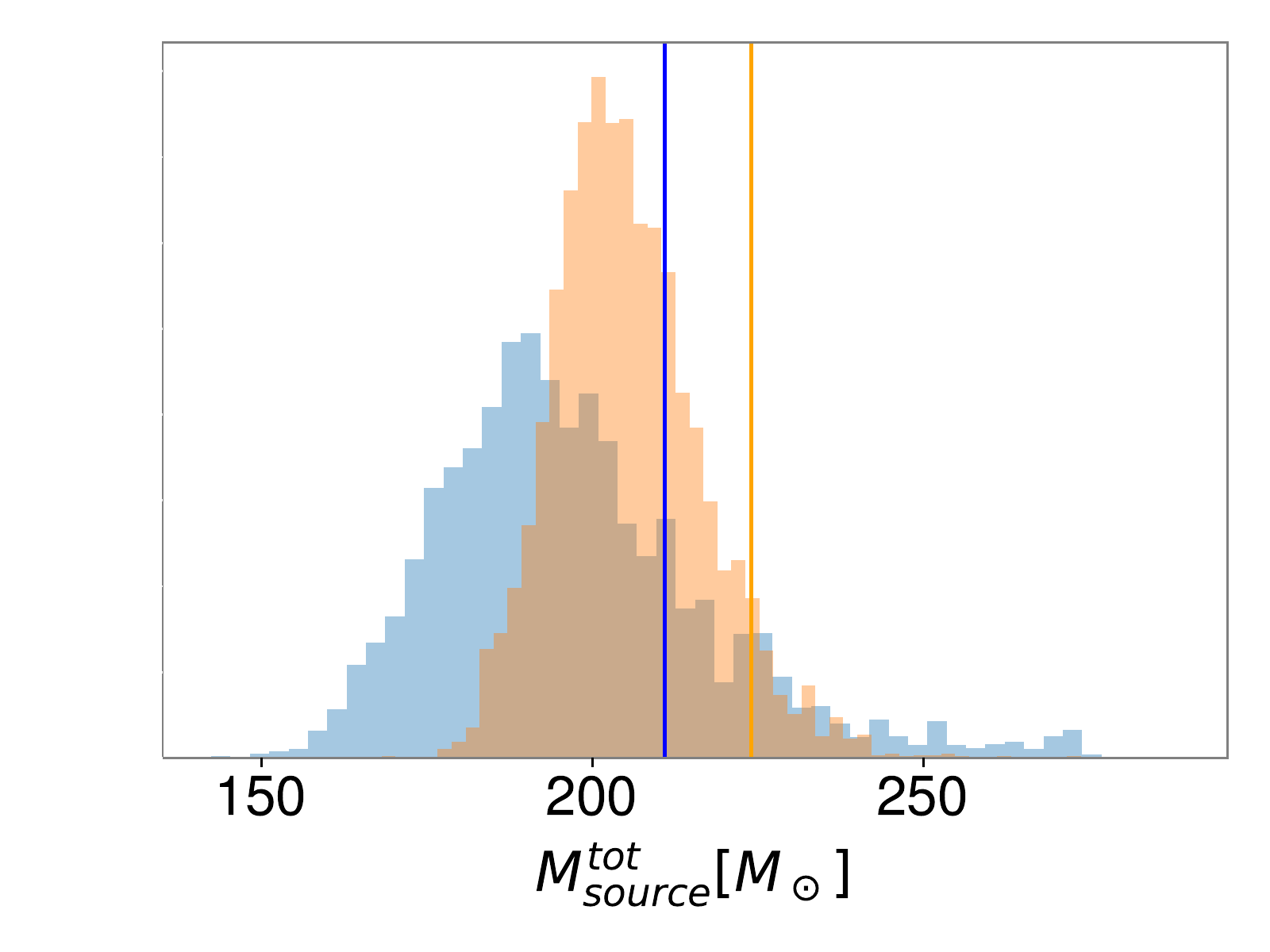}
\includegraphics[width=0.233\textwidth]{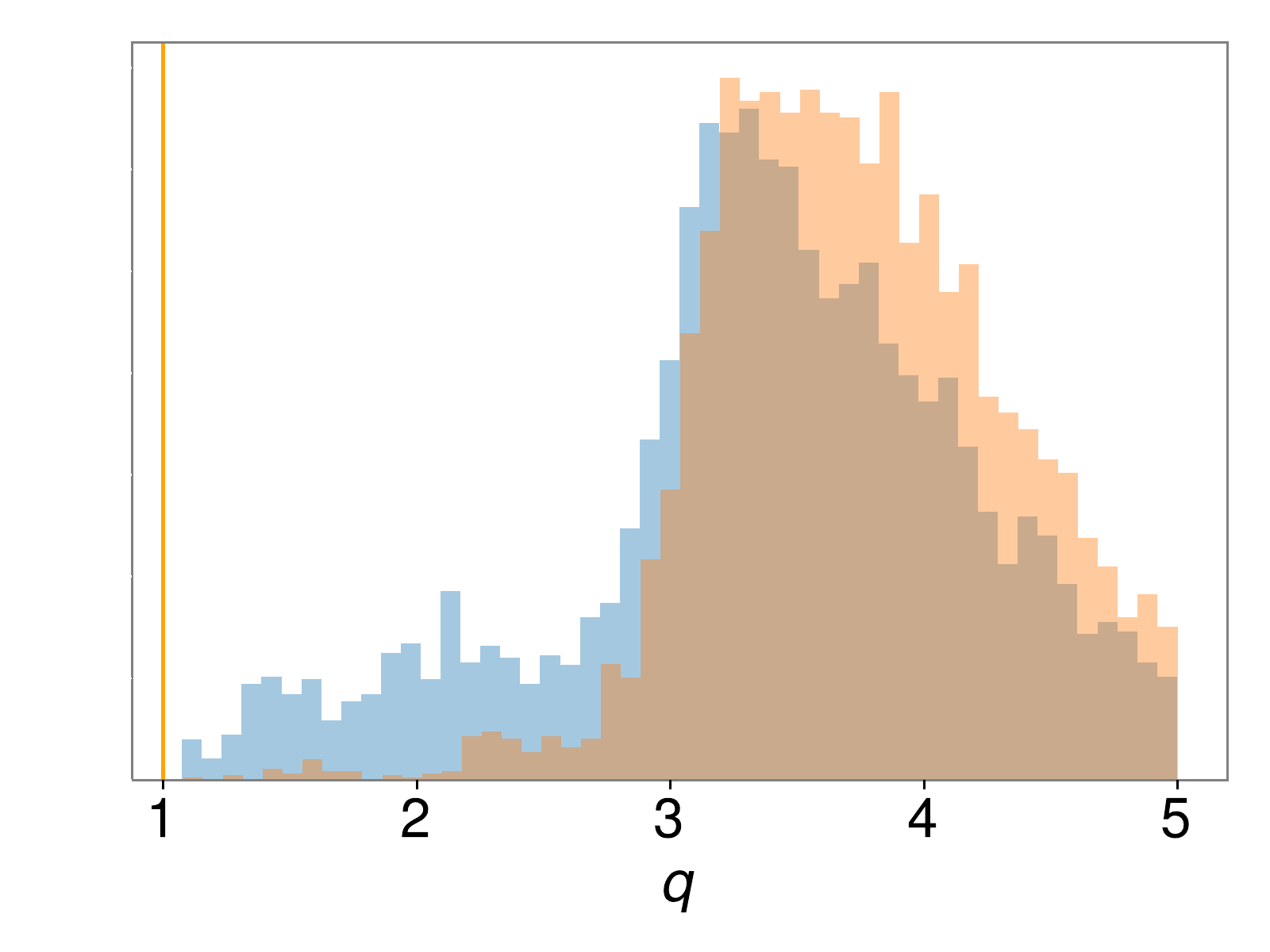}
\includegraphics[width=0.23\textwidth]{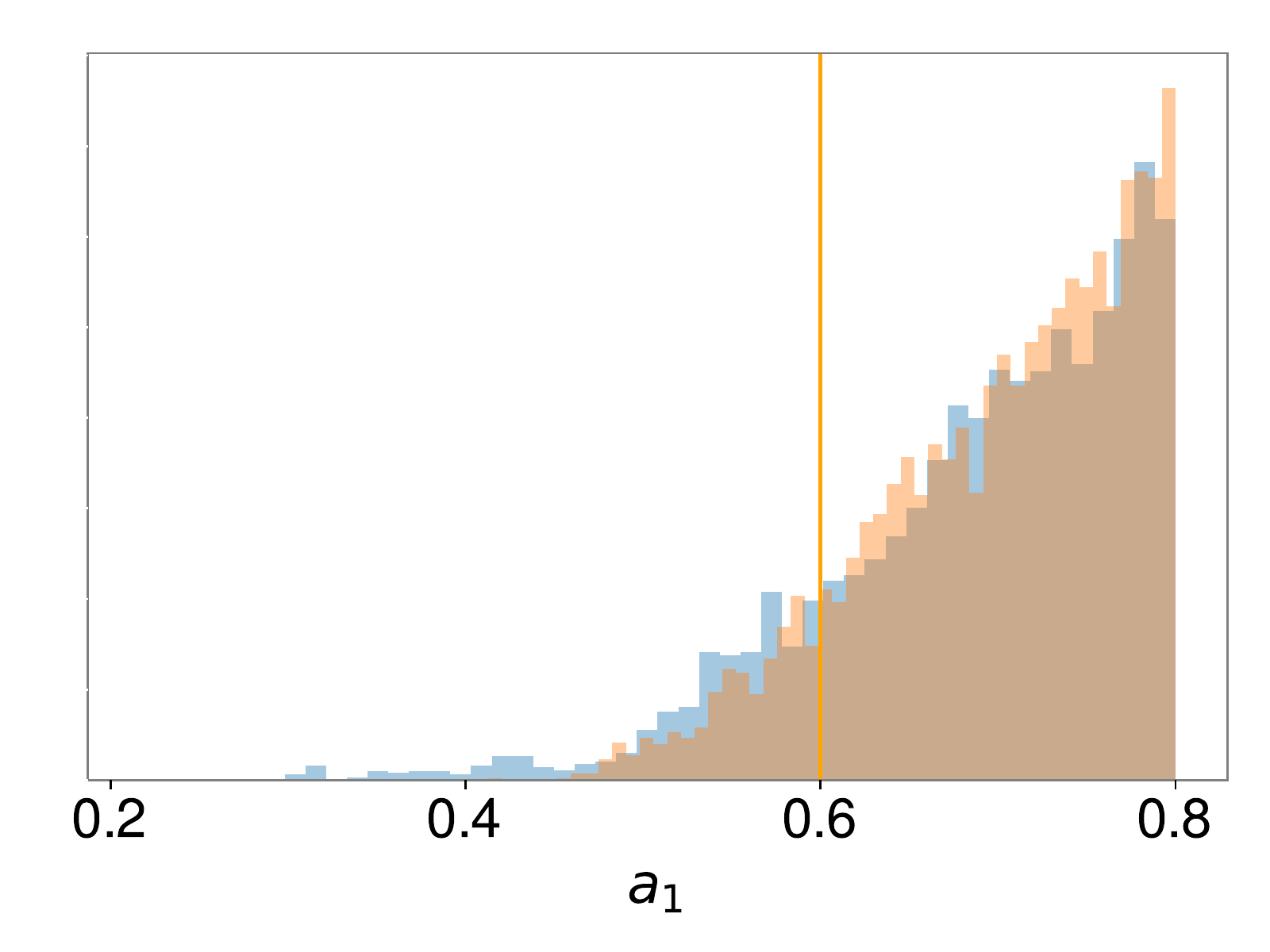}
\includegraphics[width=0.23\textwidth]{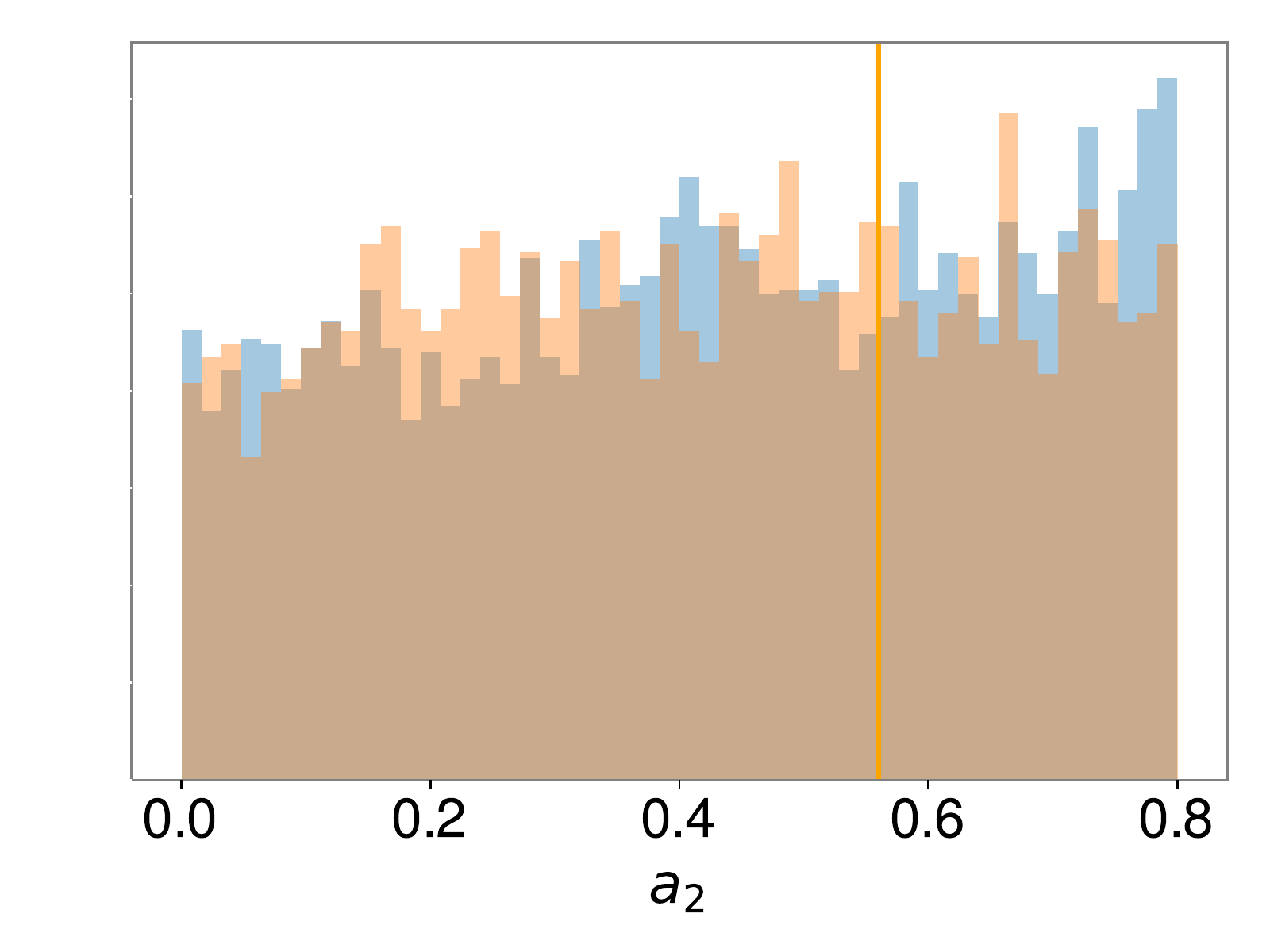}
\includegraphics[width=0.23\textwidth]{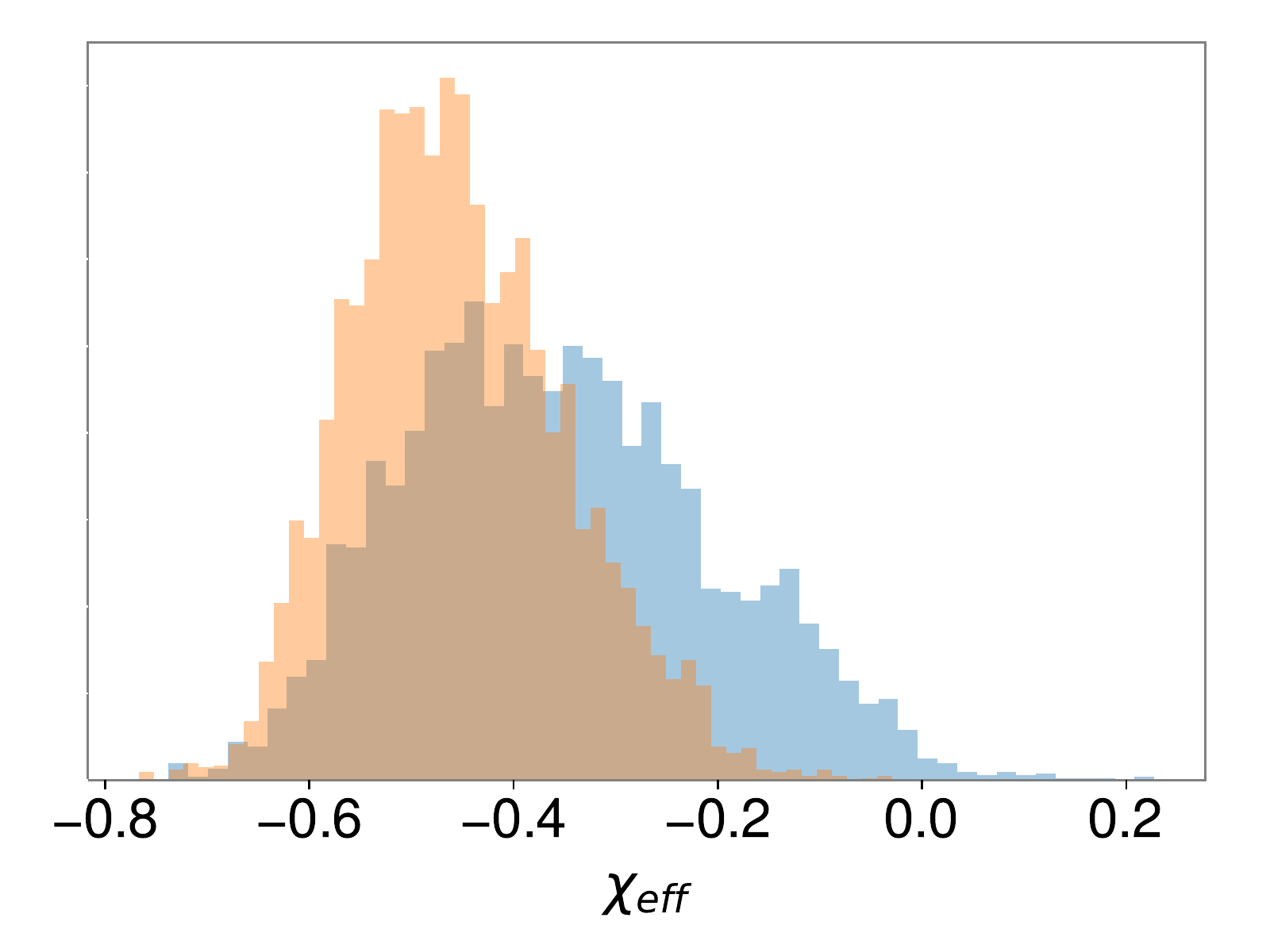}
\includegraphics[width=0.23\textwidth]{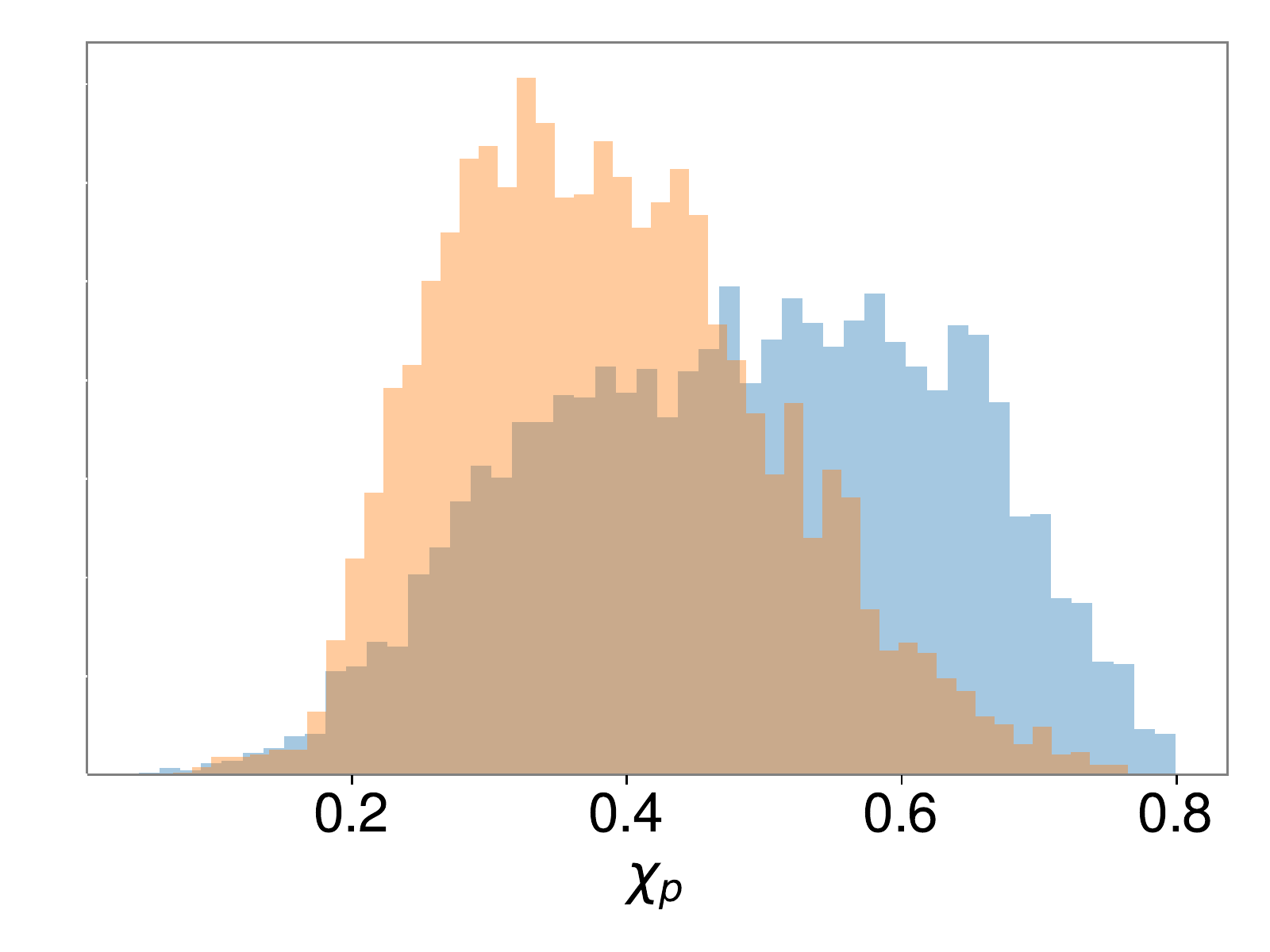}
\caption{{Parameter recovery for our spinning  $q=1$ HOC injection using the quasi-circular precessing model \tt{IMRPhenomPv2}}. 
True values are indicated by blue ($\rho_{\rm opt}=15$) and orange (25) vertical lines. 
}
\label{fig:pe_single}
\end{figure}

\paragraph*{\textbf{Results.--}}

We perform Bayesian parameter estimation and model selection using the software {\tt Bilby} \cite{Ashton:2018jfp,RomeroShaw_bilby} on the numerically simulated signals from the HOC sources reported in Table I. We scale these to total-masses in the range $M\in [100,700]\,M_\odot$. We analyse these signals with two waveform models for precessing and non-precessing BBHs respectively known as {\tt IMRPhenomPv2}  and {\tt IMRPhenomD} \cite{Hannam:2013oca,Schmidt:2012rh,Khan:2015jqa,Husa:2015iqa}. We place our HOC sources face-on (with the final-spin parallel to the line-of-sight), at two distances leading to a signal loudness (commonly denoted 
``optimal SNR'', $\rho_{\rm opt}$) typical among LIGO-Virgo observations, namely $\rho_{\text{opt}}=15$ and $25$. We consider a single Advanced LIGO detector working at design sensitivity \cite{advLIGOcurves} with a lower frequency cutoff of 20 Hz. We sample the parameter space fixing the sky-location to the true one and placing standard priors on all other parameters. For an in-depth description of the analysis setup please see the Supplementary Material.


Figure \ref{fig:spectra} shows the spectra of selected $q=2$ HOC injections together with the best fitting waveforms returned by the aligned-spin and precessing BBH models.
The shaded regions of Figure~\ref{fig:logB} \textcolor{black}{show the mass ranges where our $q=1$ HOCs can be mistaken by an apparently precessing BBH. First,} solid lines show the ratio of the Bayesian (natural)log-evidence for precessing vs. aligned-spin,  $\log\beta^{\tt IMRPhenomP}_{\tt IMRPhenomD}$, as a function of the detector-frame total mass of the HOC and the optimal SNR of the signal. Values $\log\beta^{\tt IMRPhenomP}_{\tt IMRPhenomD}>5$ indicate a strong preference for a precessing BBH. \textcolor{black}{Second, dotted lines} show the fraction of SNR that the BBH models can recover from the HOC signal, or Fitting Factor (FF) \cite{Apostolatos:1995pj}. FFs near unity indicate that the HOC signal would be perfectly fitted. Conversely, a low FF indicates that the BBH waveforms cannot fit the HOC signal, leaving significant signal residuals that would allow to discard the BBH nature of the source. \textcolor{black}{Finally, using the Akaike Information Criterion (AIC) \cite{Akaike1974,Gayathri_21g} we can estimate $\log\beta^{\tt HOC}_{\tt{IMRPhenomP}}$ as ${\text{AIC}}/2=\rho_{\rm opt}^2(1-FF^2)/2$, shown in dashed lines. Therefore, we delimit the shaded regions of Fig. 2 by $\log\beta^{\tt IMRPhenomP}_{\tt IMRPhenomD}>5$ and $\log\beta^{\tt HOC}_{\tt IMRPhenomP}<5$.} 

At the low-mass end (see Fig.~\ref{fig:spectra}, left panels) when the full HOC signal is in band, none of the BBH models fits the data well, yielding \textcolor{black}{poor} FFs and $\log\beta^{\tt HOC}_{\tt IMRPhenomP}>5$. For {\tt IMRPhenomD}, the presence of strong inspiral cycles in the detector band prevents to mimic the absence of cycles in the HOC case. Similarly, for {\tt IMRPhenomPv2}, the suppression of the inspiral right before merger is preceded by unsuppressed cycles visible in band. 

At the high-mass end (Fig.~\ref{fig:spectra}, right panels) only the late ringdown HOC signal is in band, both BBH models fit the injection with very high precision $(\text{FF} \simeq 1)$ and no preference for any model is observed ($\log\beta < 5$). 

As the source mass decreases, details of the early ringdown and merger signal that contain information about the HOC origin of the remnant become visible in band. While {\tt IMRPhenomD} cannot mimic these details (see the fast decrease of its FF) {\tt IMRPhenomPv2} fits the HOC signal perfectly down to $300\,M_\odot$,  i.e.~with FF=1. This leads to an increasing preference for {\tt IMRPhenomPv2} for decreasing mass and increasing signal loudness. 

For intermediate masses we find a mass range in which {\tt IMRPhenomPv2} large FFs while {\tt IMRPhenomD} yields much lower values. The reason is that the strong modulation induced by precession suppresses the inspiral cycles before merger while, unlike for lower masses, the earlier inspiral lays out of band, perfectly mimicking the absence of inspiral cycles that characterises HOCs (Fig.~\ref{fig:spectra}, central panels). 
%
%
This causes a strong preference for a precessing vs.~aligned-spin BBH model\textcolor{black}{, indicated by $\log\beta^{\tt IMRPhenomP}_{\tt IMRPhenomD}>5$, accompanied by no strong preference for the HOC model, $\log\beta^{\tt HOC}_{\tt IMRPhenomP}<5$. For SNR $\simeq 15 (25)$ this happens when the total source-frame mass is within $M\in (125,300)M_\odot$  $((200,440)M_\odot$)} (orange and blue regions of Fig.~2).

\begin{figure*}
\includegraphics[width=0.32\textwidth]{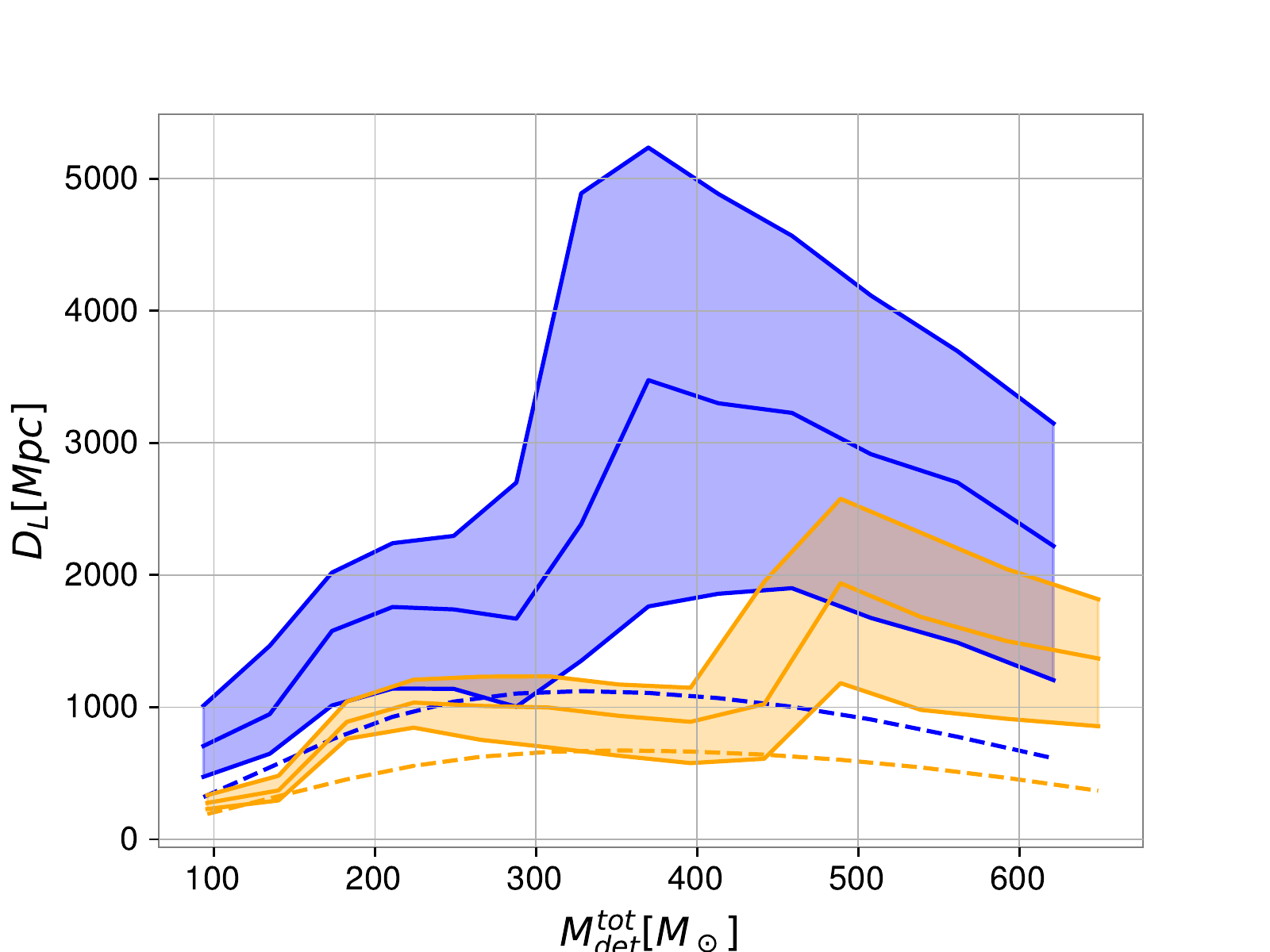}
\includegraphics[width=0.32\textwidth]{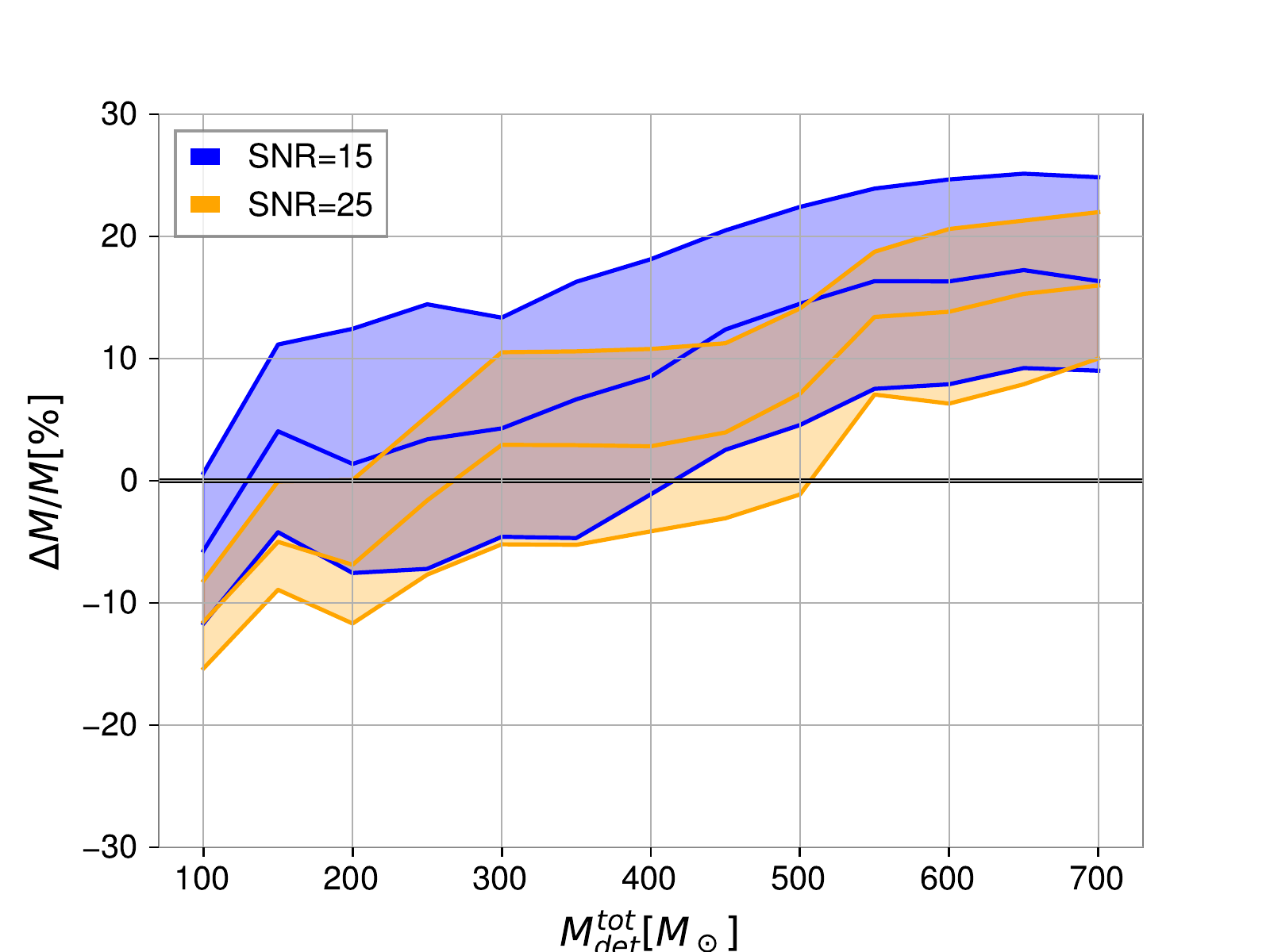}
\includegraphics[width=0.32\textwidth]{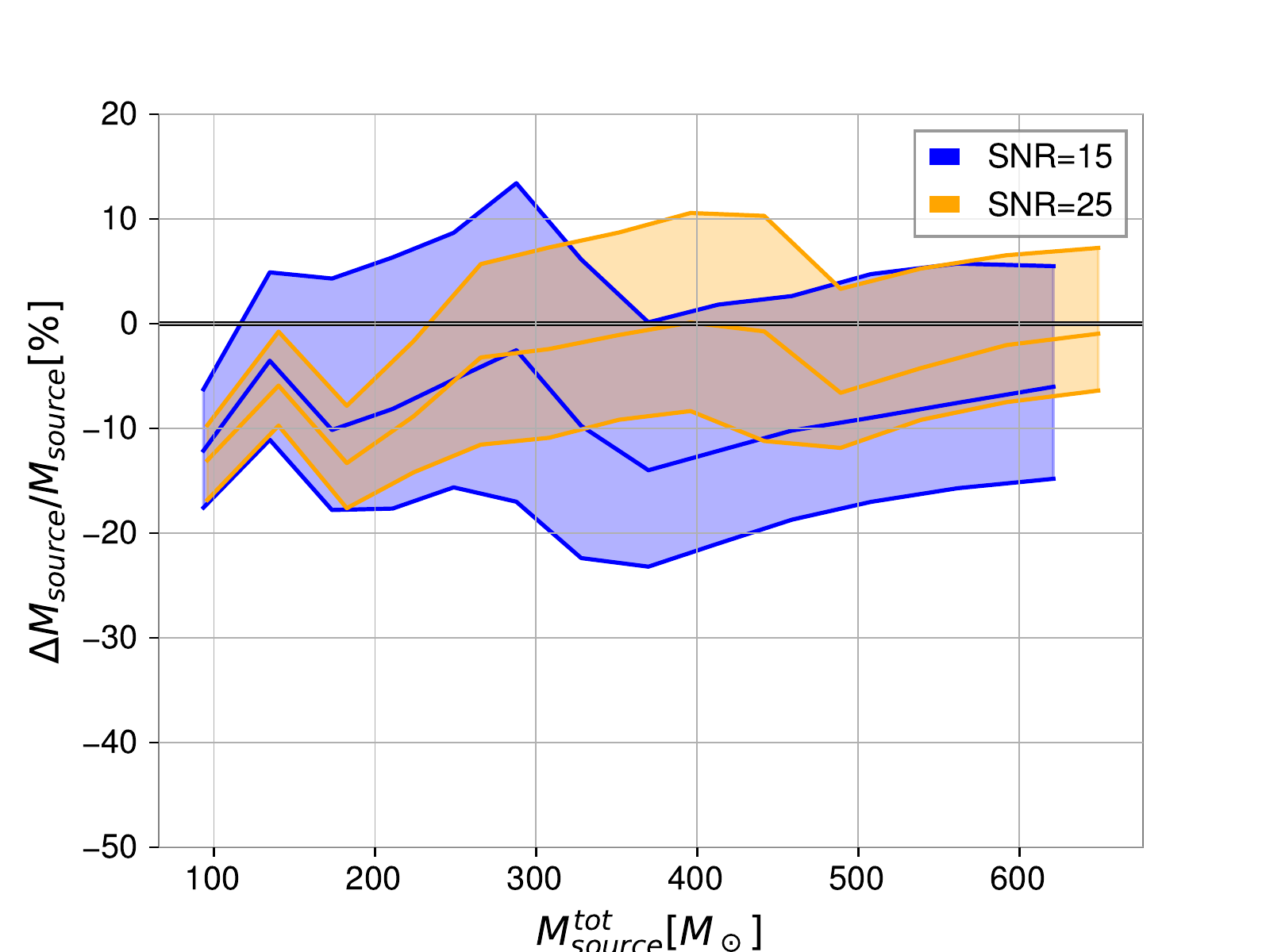}
\includegraphics[width=0.32\textwidth]{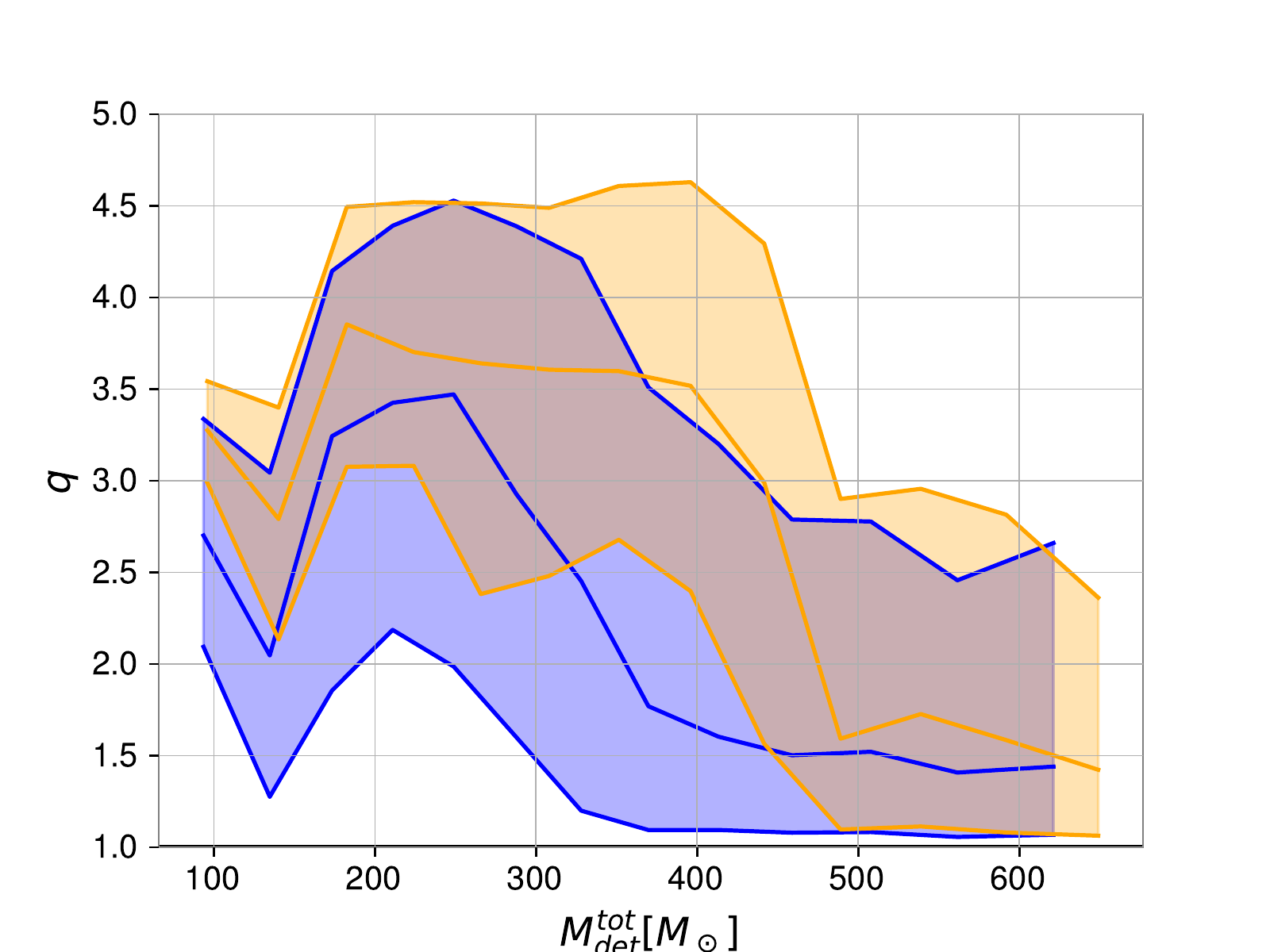}
\includegraphics[width=0.32\textwidth]{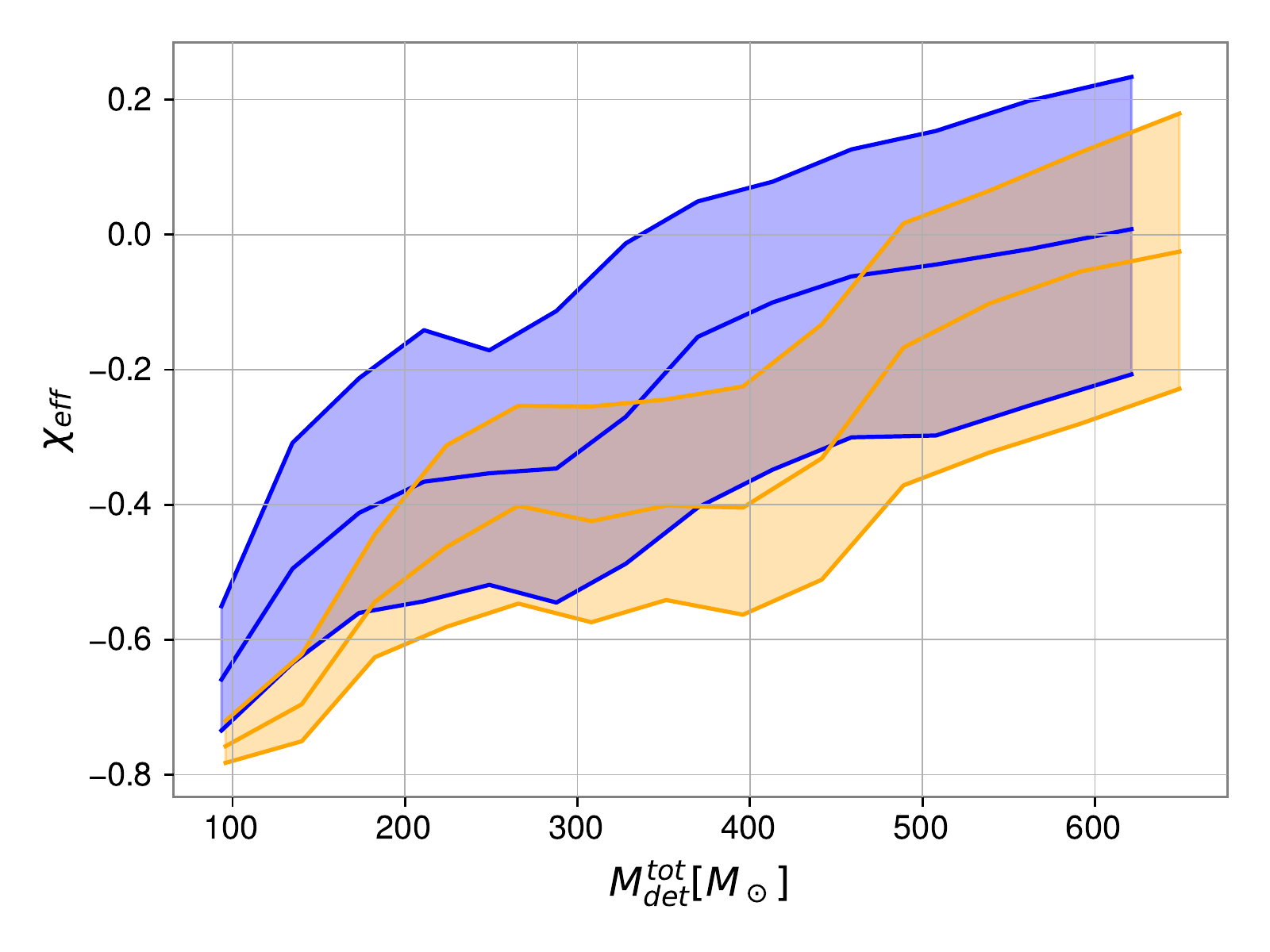}
\includegraphics[width=0.32\textwidth]{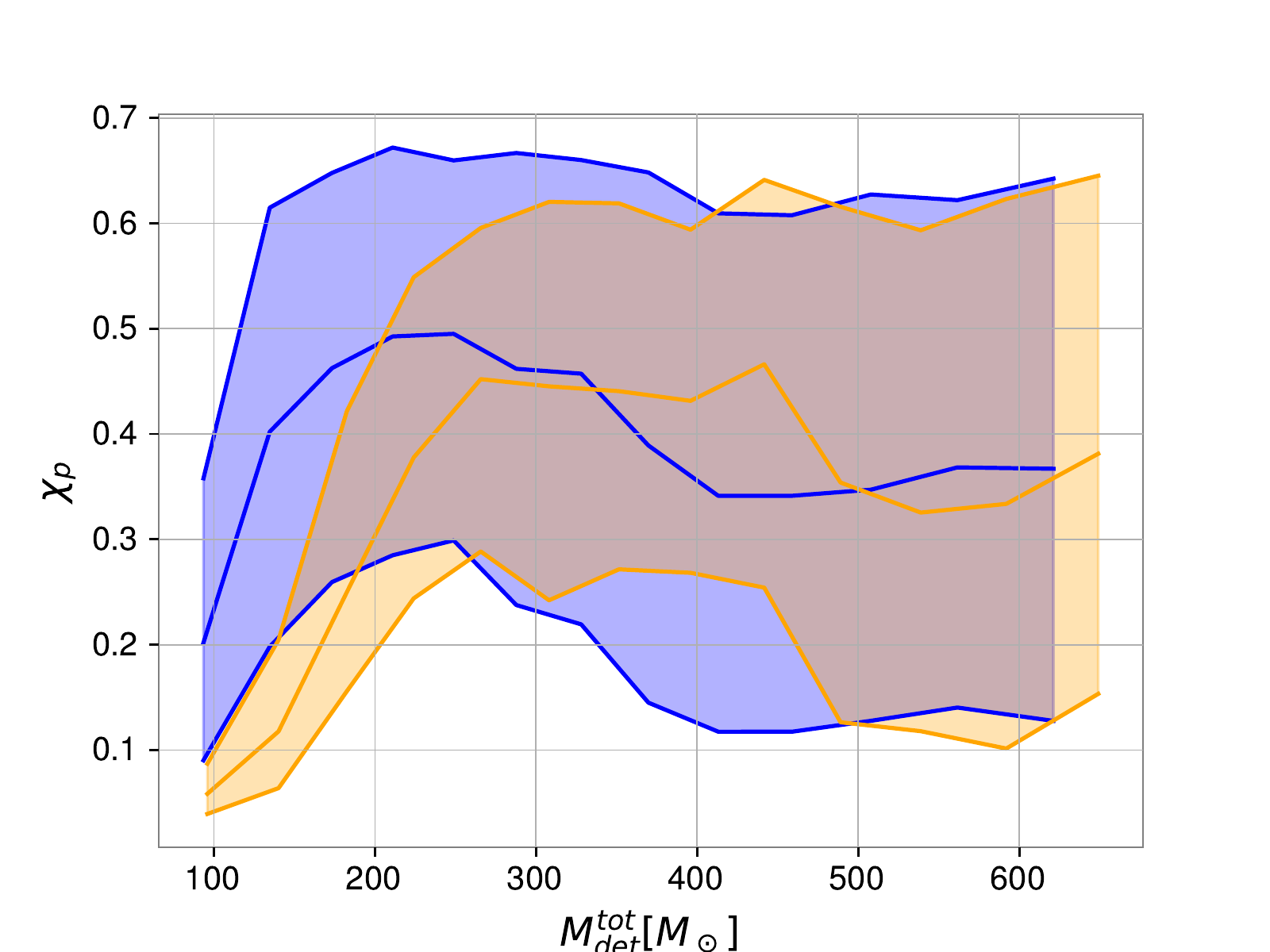}
\caption{Parameter recovery for our spinning $q=1$ HOC injection, varying the total mass, and using the {\tt IMRPhenomPv2} model for $\rho_{\rm opt}=15$ (blue) and $25$ (orange). The shaded contours represent $90\%$ confidence intervals. For the luminosity distance the dashed lines represent the true values of the injection. 
}
\label{fig:pe}
\end{figure*}

Figure ~\ref{fig:pe_single} shows posterior parameter distributions for our spinning equal-mass HOC in Table I, analysed with {\tt IMRPhenomPv2}. We consider $\rho_{\rm opt}=15$ (blue) and 25 (orange) and a total mass of $250\,M_\odot$ in the detector-frame. The BBH model tries to mimic the HOC signal by \textcolor{black}{preferring} a strongly precessing BBH (as indicated by the spin-precession parameter $\chi_{\rm p}$, zero for non-precessing systems \cite{Schmidt:2014iyl,Hannam:2013oca}). The posterior distribution of the primary spin hits the upper prior boundary. In addition, because BBH mergers are much \textcolor{black}{louder} than HOCs, the BBH model overestimates the source distance. This, together with fair estimates of the detector-frame mass $M_{\rm det}$, leads to underestimated source-frame masses, $M_{\rm source}=M_{\rm det}/(1+z)$. Moreover, the large mass ratio $q$ and the negative effective-spin $\chi_{\rm eff}$ \footnote{$\chi_{\rm eff}=\frac{m_1\chi_1^{\parallel}+m_2\chi_2^{\parallel}}{m_1+m_2}$, with $\chi^{\parallel}_i$ being the component of the $i$-th spin along the orbital angular momentum}, indicate that the BBH model tries to shorten the waveform as much as possible, to leave the 
low-frequency inspiral out of the detector band. The negative $\chi_{\rm eff}$ is also a consequence of the low final spin of our binaries (see Table I) which the BBH model \textcolor{black}{reproduces} by anti-aligning the spins with the orbital angular momentum.

Figure ~\ref{fig:pe} shows our posterior $90\%$ confidence intervals on several parameters for cases with source-frame masses $\sim(100-650)M_\odot$. At low masses, results are consistent with the ones previously described. For the largest masses, when only the ringdown is in band, the estimated detector-frame total mass is slightly overestimated. The reason is that the BBH model tries to obtain the same remnant BH mass as the HOC. Since BBHs are more luminous, this requires a larger initial mass. \textcolor{black}{Interestingly, the distance and detector-frame mass estimates combine to yield non-biased estimates of the detector-frame mass, despite a slight deviation of the median estimates toward lower values}. Together with a bias to large mass ratios, this leads to wrong estimates of the individual masses, whose implications we discuss next. These findings hold for all HOC models in Table I.\\

\begin{figure*}
\includegraphics[width=0.49\textwidth]{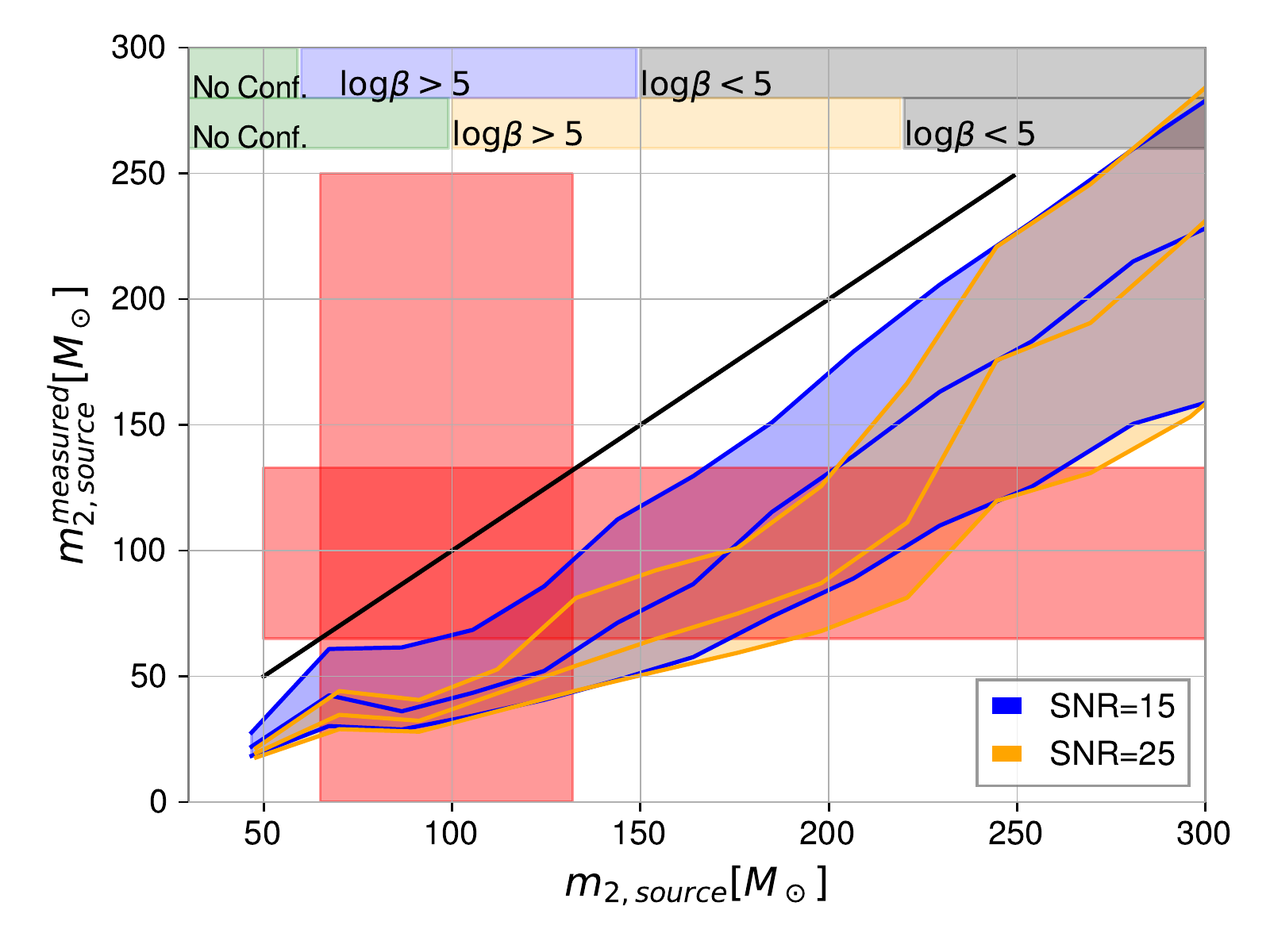}
\includegraphics[width=0.49\textwidth]{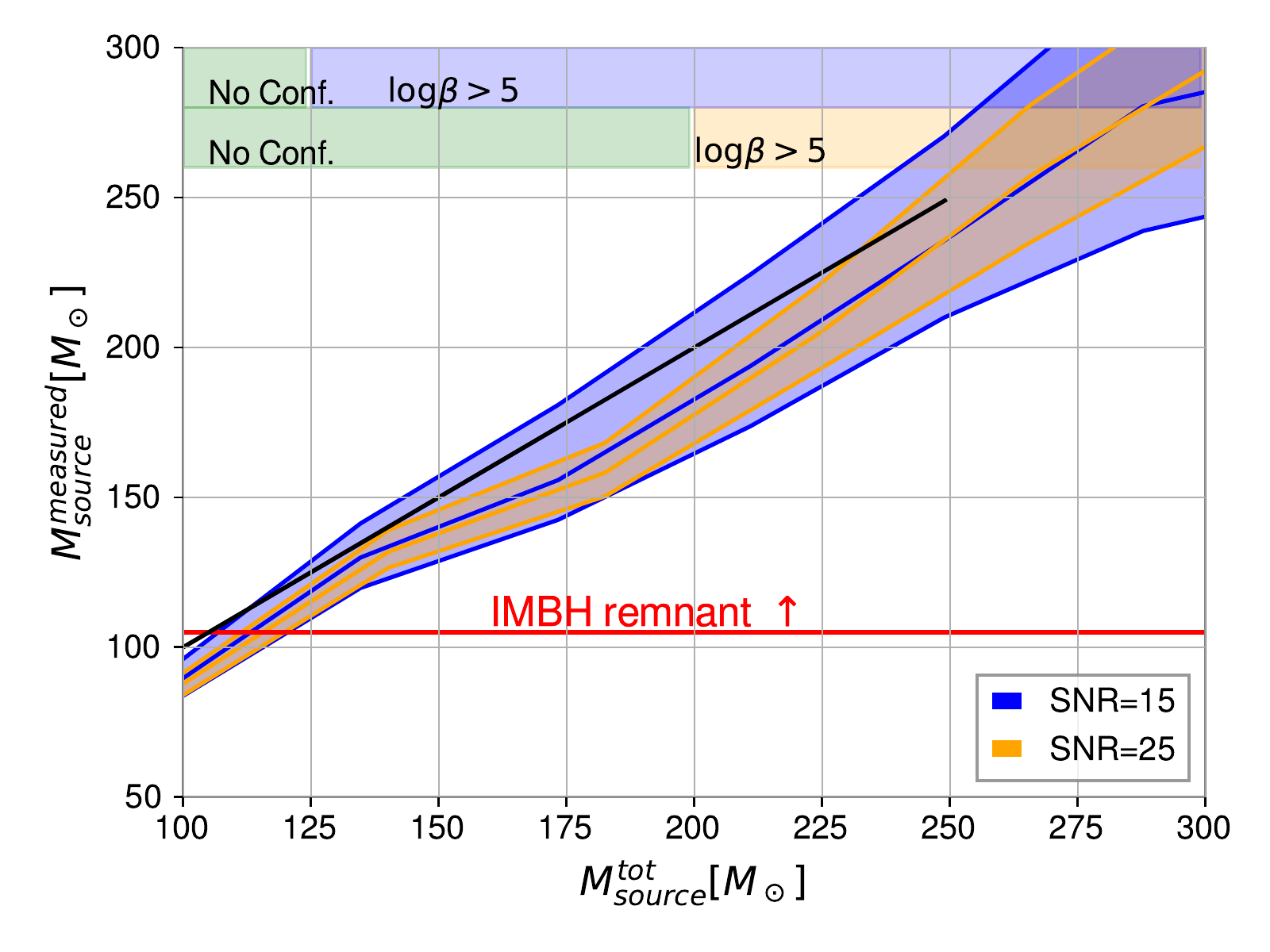}
\caption{
$90\%$ confidence intervals for the secondary mass (left) and total mass (right) of an equal-mass HOC, measured in the source-frame. The diagonal black line denotes ``zero bias''. Red regions enclose the boundaries of the PISN gap. The top blue and orange regions show the mass ranges in which the source may be confused with a precessing BBH, for $\rho_{\rm opt}=15$ and 25, respectively. Top green regions denote ranges in which a BBH nature of the source can be discarded.  Grey regions denote ranges in which no model is preferred. 
}
\label{fig:gap}
\end{figure*}


\paragraph*{\textbf{Astrophysical implications.--}}

Figure ~\ref{fig:gap} shows $90\%$ confidence intervals for the total mass and secondary source-frame mass of an equal-mass HOC, using {\tt IMRPhenomPv2}. Both are biased to lower values which has two main consequences.

First, for $\rho_{\rm opt}=15$, HOCs with \textcolor{black}{$ M_{\rm source} < 120 M_\odot$} would have non-zero support for total measured masses below $100\,M_\odot$, 
preventing a conclusive detection of a remnant IMBH. However, in this range, BBH waveforms would not fit HOC ones (see Figs.~1 and 2) making the BBH nature of the source easy to discard.
Therefore, the measurement of an IMBH is robust against the bias we describe.

Second, for certain component-mass range outside (inside) the PISN gap, precessing waveforms may report secondary mass estimates inside (outside) the gap. In particular, for an equal-mass HOC with \textcolor{black}{$m_{\rm 1,source}=m_{\rm 2,source} \in (130,200)\,M_\odot$} (above the gap) we recover posterior distributions for $m_2$ with strong support within the PISN gap. Since \textcolor{black}{in this region $\log\beta^{\tt HOC}_{\tt{IMRPhenomP}}<5$}, one cannot discard the BBH nature of the source in a straightforward way. Yet, a strong preference for precession shall be flagged as a symptom that the source may be a HOC with masses outside the gap.  
For larger masses, for which such preference would not occur (grey regions in Fig.~\ref{fig:gap}) no such diagnose would be possible. Finally, when the source masses are in the interval $(65,130)\,M_\odot$ (within the PISN gap),  {\tt IMRPhenomPv2} reports a secondary mass with no support within the gap.

\paragraph*{\textbf{Conclusions.--}}
GW signals from head-on BH collisions (HOCs) can be confused with those emitted by precessing quasi-circular BBH mergers. For a SNR=15, similar to that of GW190521,
this confusion happens for HOC masses \textcolor{black}{$M\in (125,300)M_\odot$}. In this interval, the absence of inspiral cycles  characteristic of HOCs can be mimicked by the suppression of signal power before merger induced by orbital precession. Therefore, discerning between head-on and precessing scenarios would only be possible via model selection with currently unavailable models for HOCs and the placement of appropriate priors on the astrophysical probability of such collisions. The confusion brings significant biases in the source parameters. HOCs in the IMBH range can be mistaken for highly precessing IMBH ones at a much larger distance, causing an underestimation of the source-frame mass by $10$--$20\%$, consequently biasing the component masses. HOCs with component masses outside (inside) the PISN mass gap may be misinterpreted as BBHs with a component mass inside (outside) the gap, leading to a fake violation (compliance) of the PISN gap. 

The recent GW observation GW190521 \cite{GW190521D,GW190521I} is a short signal with signatures of precession and a component BH in the PISN gap. Our results suggest that it may admit a head-on (or extremely eccentric) BBH interpretation, \textcolor{black}{as shown by \cite{Gayathri_21g,Romero-Shaw:2019itr,Proca} after the release of this work}. However, HOCs are not only much rarer than quasi-circular mergers but would be further disfavoured if an uniform distribution of sources in the Universe is assumed, due to their intrinsic weakness. Additionally, the remnant of GW150921 has a spin $a_{\rm f} = 0.72^{+0.10}_{-0.11}$, very difficult to obtain through HOCs of low mass-ratio. Nevertheless, given the recent claim of an associated electromagnetic counterpart to GW190521 located at a much shorter distance than estimated by quasi-circular BBH models \cite{Graham2020}, our results suggest that a highly eccentric merger (if not necessarily head-on) may reconcile both distance estimates, \textcolor{black}{as later shown by \cite{Gayathri_21g}}.

\textcolor{black}{Finally, in \cite{Proca} we analyse in detail public GW190521 data \cite{GWOSC} in the HOC scenario, showing that while being only marginally consistent with black-hole HOC, it is highly consistent with that of horizonless objects known as Proca stars \cite{brito2016proca}}.\\


\paragraph*{\textbf{Acknowledgements.}}
We thank Paul D. Lasky, Simon Stevenson,  Nelson Christensen for comments on the manuscript and Tito dal Canton and Mark Gieles for useful comments. We also thank the anonymous referees for valuable comments. JCB is supported by the Australian Research Council Discovery Project DP180103155 and the Direct Grant, Project 4053406, from the Research Committee of the Chinese University of Hong Kong.  NSG is supported by the Funda\c c\~ao para a Ci\^encia e a Tecnologia (FCT) projects PTDC/FIS-OUT/28407/2017, UID/FIS/00099/2020 (CENTRA), and CERN/FIS-PAR/0027/2019. The project that gave rise to these results also received the support of a fellowship from ”la Caixa” Foundation (ID
100010434) and from the European Union’s Horizon
2020 research and innovation programme under the
Marie Skłodowska-Curie grant agreement No 847648.
The fellowship code is LCF/BQ/PI20/11760016.
JAF is supported by the Spanish Agencia Estatal de Investigaci\'on  (PGC2018-095984-B-I00) and  by the  Generalitat  Valenciana  (PROMETEO/2019/071). 
This work has further been supported by  the  European  Union's  Horizon  2020  research  and  innovation
(RISE) programme H2020-MSCA-RISE-2017 Grant No.~FunFiCO-777740. Parameter estimation runs were done in the CIT Caltech cluster. 


\section*{Supplementary Material}

\paragraph*{\textbf{Analysis Details.}}

\begin{figure}[ht!]
\centering
\includegraphics[width=0.42\textwidth]{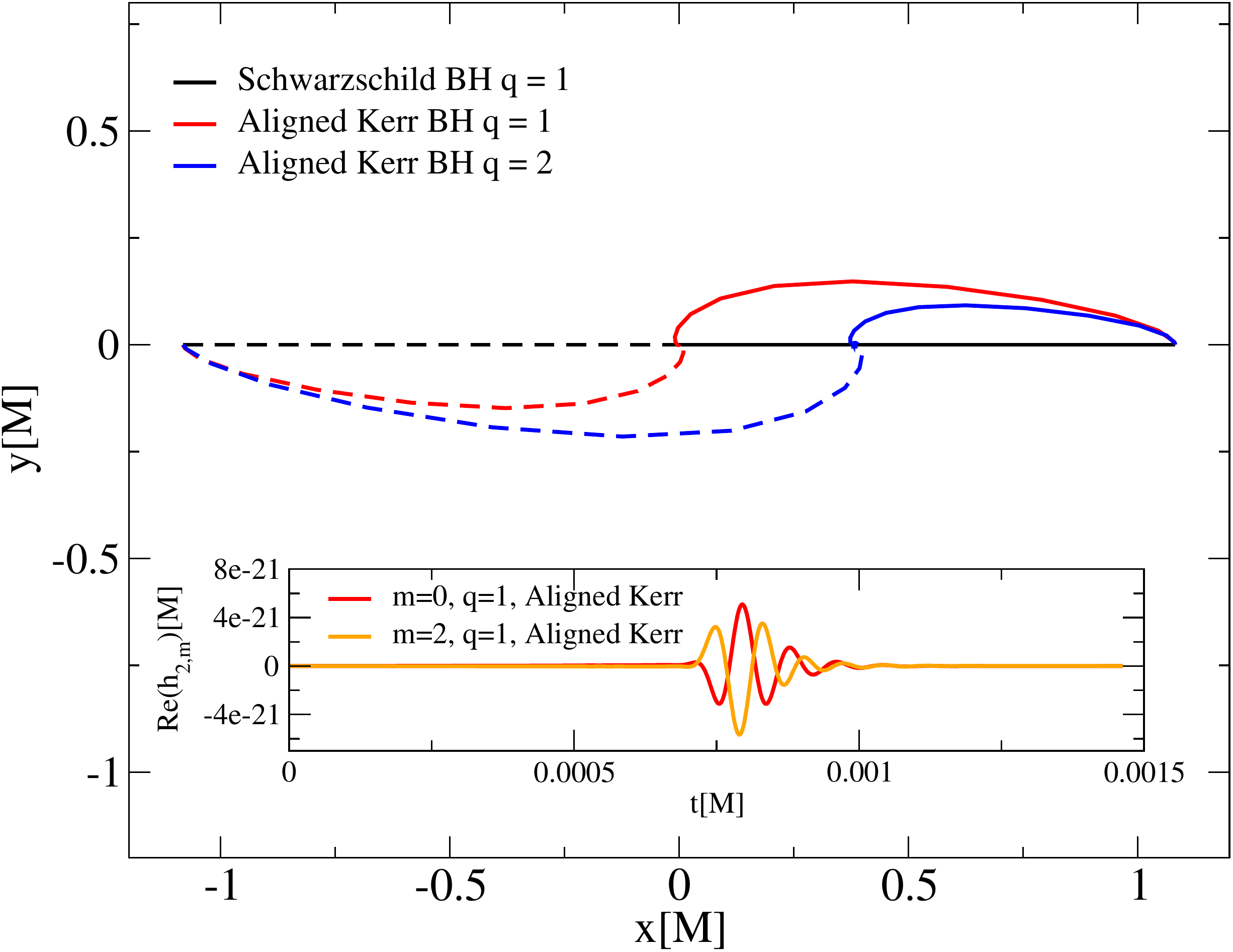}
\\
\hspace{-0.6cm}
\includegraphics[width=0.48\textwidth]{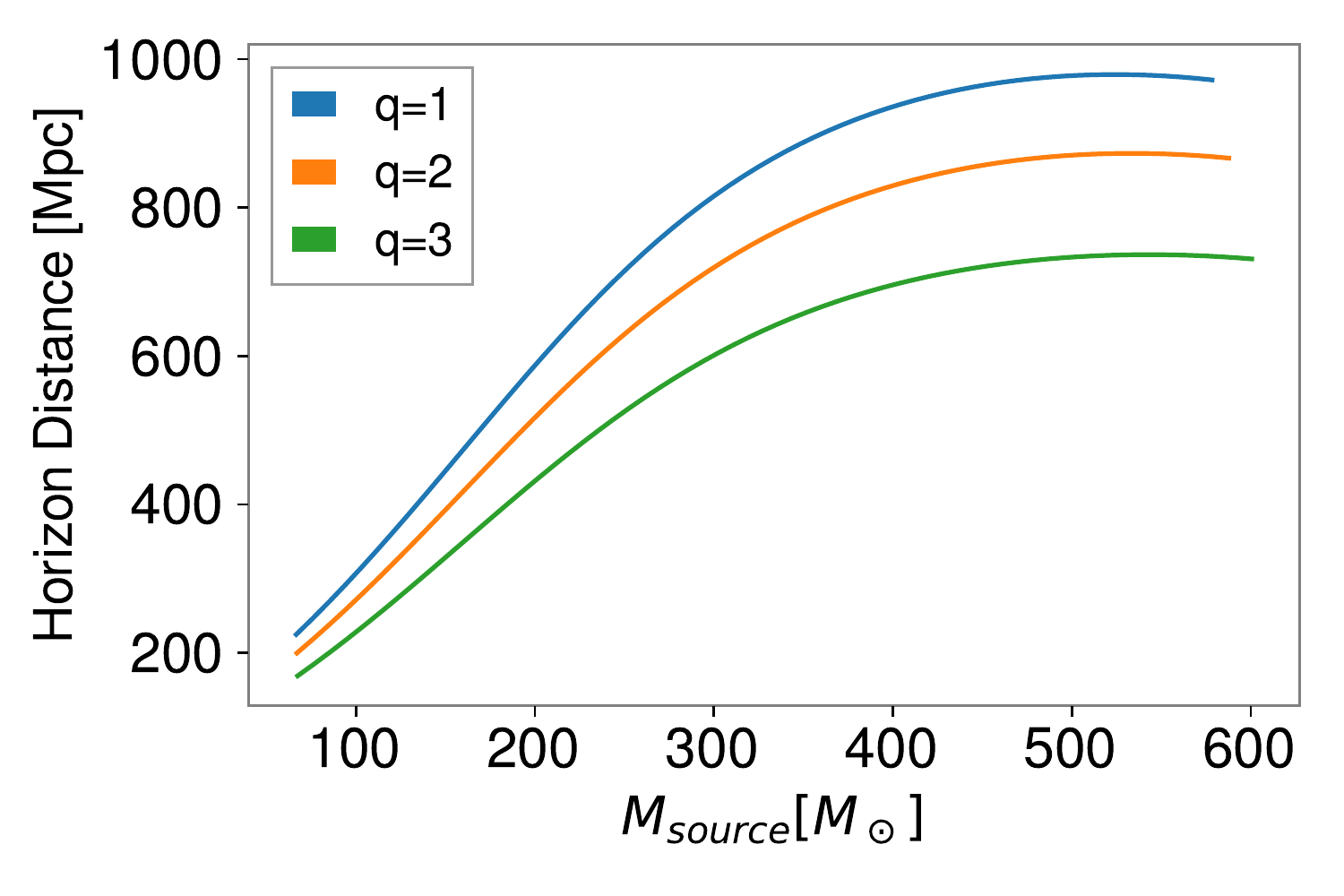}
\caption{Top: HOC trajectories for two Schwarzschild BHs and two Kerr BHs with different mass ratio and aligned spins $a_1$ and $a_2$ (see Table 1 in the main text). The BH spins induce a frame-dragging effect, curving the trajectories. Bottom: Average distance at which a HOC can produce a SNR of 12, using a triple detector network (LIGO Hanford-Livingston and Virgo) at design sensitivities.
}
\label{fig:trajectories}
\end{figure}

We perform parameter estimation on a family of numerically simulated signals from HOCs using two phenomenological waveform models for the inspiral-merger-ringdown of precessing and aligned-spin BBHs, respectively known as {\tt IMRPhenomPv2} 
and {\tt IMRPhenomD} \cite{Hannam:2013oca,Schmidt:2012rh,Khan:2015jqa,Husa:2015iqa} \footnote{We also performed another analysis using the model NRSur7dq4 \cite{NRSur7dq4}. However, this is limited to mass ratios $q \leq 4$, which caused our posteriors to rail against this limit.} and the parameter-estimation code {\tt Bilby} \cite{Ashton:2018jfp,RomeroShaw_bilby}. The numerical simulations of the HOCs and the GW extraction are performed with the \texttt{Einstein Toolkit}~\cite{Loffler:2011ay,EinsteinToolkit,zilhao2013introduction} using the \texttt{Cactus} framework with mesh refinement.  
The HOC initial data (see Table I in the main text) are obtained from the \texttt{TwoPunctures} thorn~\cite{Ansorg:2004ds} and we use the \texttt{MacLachlan} code~\cite{Brown:2008sb} to solve Einstein's equations. Examples of the HOC trajectories and an illustrative GW strain signal are shown in the top panel of Fig.~\ref{fig:trajectories}.\\

 Our HOCs cover the total redshifted (detector-frame) mass range $M\in[100,700]\,M_\odot$ related to the source-frame mass by $M = (1+z) M_{\rm source}$, with $z$ the redshift. While the BBH emission is dominated by the quadrupole $(\ell,m)=(2,\pm 2)$ modes \cite{Bustillo:2016gid,Capano:2013raa,CalderonBustillo:2017skv},
 both the $(\ell,m)=(2,\pm 2)$ and $(\ell,m)=(2,\pm 0)$ are co-dominant for HOCs (see Fig. \ref{fig:trajectories}). We choose a face-on source so that the HOC emission only includes the $(2,2)$ mode, minimising waveform systematics.

We inject simulated signals from HOCs in zero noise. We use the standard likelihood for GW transients \cite{Finn1992,Romano2017}
\begin{equation}
   \log{\cal{L}}(\theta|d)=-(d-h(\theta)|d-h(\theta))/2.
\end{equation}
Here, $d$ denotes our injection, $h(\theta)$ a BBH template $h$ with source parameters $\theta$, and $(a|b)$ the noise-weighted inner product \cite{Cutler:1994ys}, 
\begin{equation}
   (a|b)= 4 \Re \int \frac{\tilde{a}(f)\tilde{b}^{*}(f)}{S_n(f)}df,
\end{equation}
where $S_n(f)$ is the one-sided power spectral density of the detector. We consider a single Advanced LIGO detector working at design sensitivity \cite{advLIGOcurves} with a lower frequency cutoff of 20Hz. We characterise the loudness of our injections by the optimal SNR, $\rho_{\rm opt}=(h|h)^{1/2}$, inversely proportional to the luminosity distance $D_L$. 

\begin{figure}[ht!]
\centering
\includegraphics[width=0.42\textwidth]{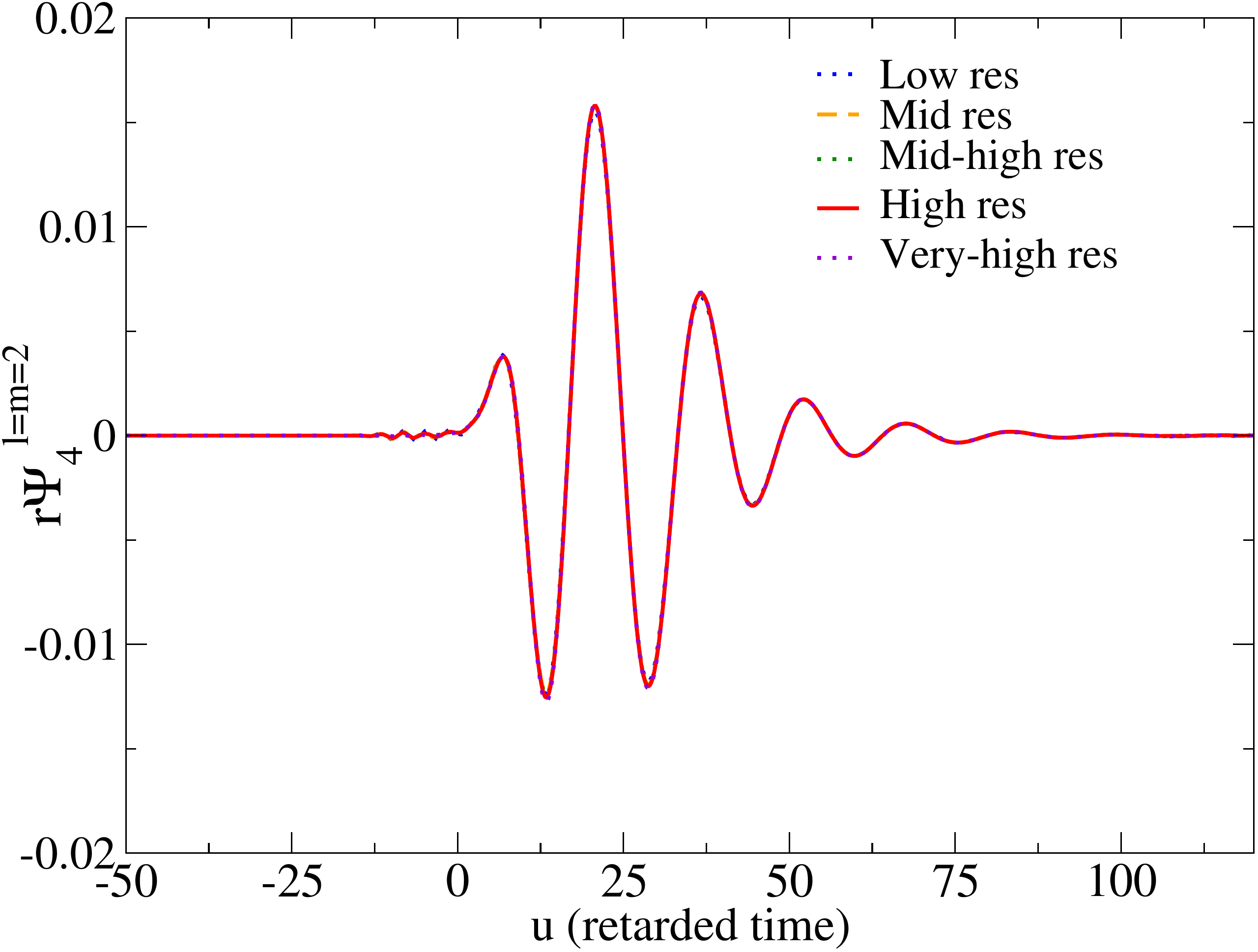}
\includegraphics[width=0.42\textwidth]{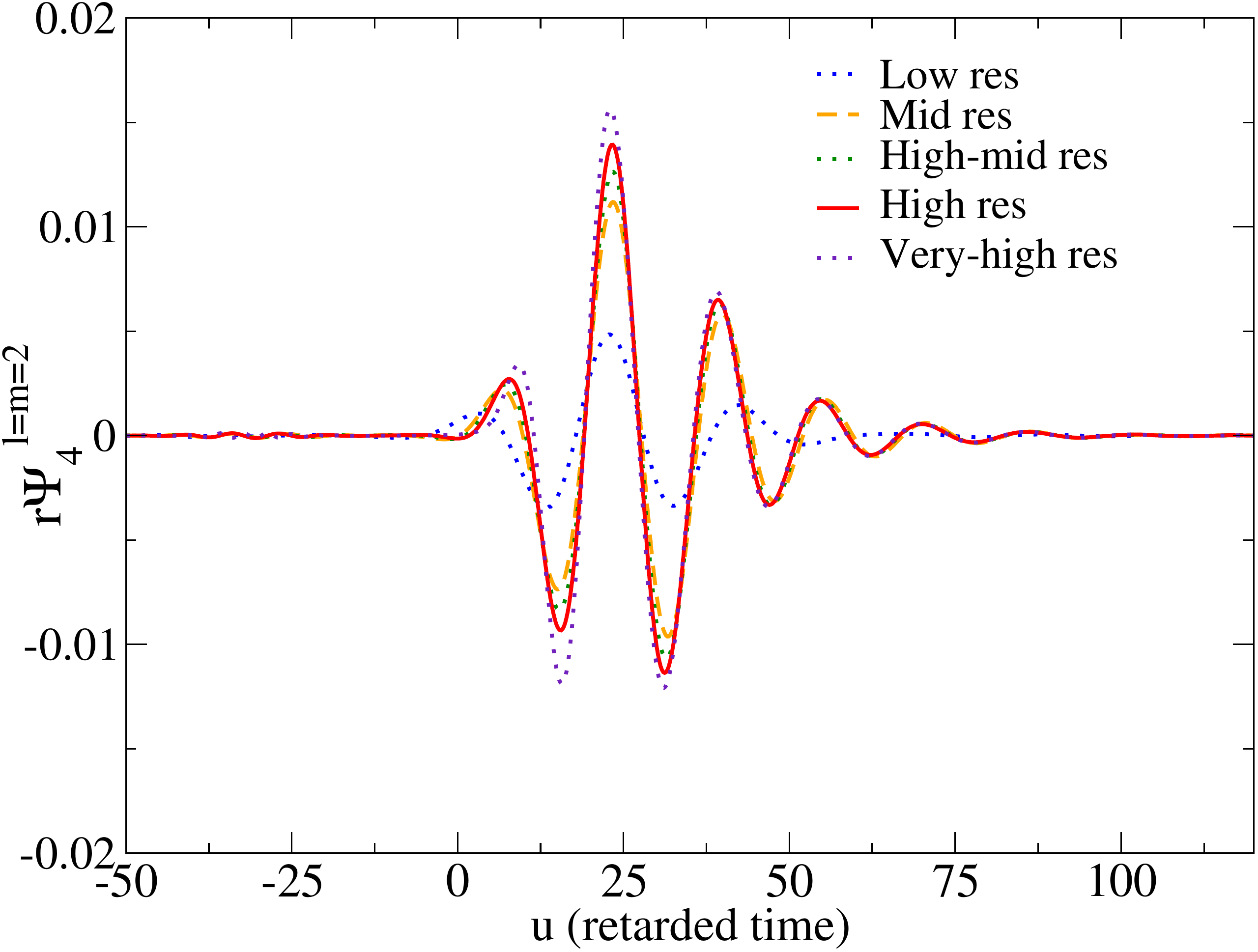}
\caption{Gravitational waveform of an equal-mass HOC extracted at $r=50M$ (top panel) and at $r=150M$ (bottom panel) with three different resolutions, expressed in terms of the Newman-Penrose scalar $\Psi_4$. We show the quadrupole $(2,2)$-mode, which dominates the face-on emission.
}
\label{fig:converg}
\end{figure}

We draw our attention to two main figures of merit: The first one is the ratio of the Bayesian evidence $B$ for our precessing and non-precessing BBH models, $\log\beta=\log B(\chi_p)-\log B(\chi_p=0)$. To compute the Bayesian evidence, we place flat priors on the detector-frame total mass and mass ratio parameters with ranges $M_{\rm tot}\in[10,1000]\,M_\odot$ and $q\in[1,5]$ and a spin prior with components uniformly distributed on the sphere. We use a prior on distance uniform in co-moving volume, a standard, isotropic prior on source orientation and fix the sky-location to the true one. The second figure of merit is the fitting factor (FF) \cite{Apostolatos:1995pj} between the HOC waveforms and our BBH waveform models. We compute the FF as the fraction of SNR that the maximum likelihood (best fitting) template can recover from our injection with, using $\rho_{\rm opt}=100$. 

A low FF indicates that the models miss an important fraction of the signal, leaving significant residuals in the data that would reveal that the signal is not a BBH. Instead, a large FF indicates that the signal is well reproduced by the BBH models and could be confused with a BBH. \textcolor{black}{To determine if such confusion is possible, we estimate the $\log\beta$ for HOC vs.~recession. In the absence of a continuous HOC waveform model that allows us to perform parameter inference, we estimate the relative Bayes Factor via the Akaike Information Criterion (AIC), given by AIC$=2\log{{\cal{L}}_{max}^{\tt HOC}}-2\log{{\cal{L}}_{max}^{\tt IMRPhenomP}}-2\Delta N = 2 \log\beta^{\tt{HOC}}_{\tt {IMRPhenomP}}$, where $\Delta N$ would denote the extra number of degrees of freedom of $\tt {IMRPhenomP}$, which we set to zero. Because $\log{{\cal{L}}_{max}}=-\rho_{opt}^2 (1-FF^2)/2$ and for HOC, $FF_{\tt HOC} = 1$, we obtain $\log\beta^{\tt{HOC}}_{\tt {IMRPhenomP}}=\rho_{\rm opt}^2(1-FF_{\tt IMRPhenomP}^2)/2$. We consier that a HOC can be confused with a Precessing BBH a if $\log\beta^{\tt{HOC}}_{\tt {BBH,Precession}} \leq 5$.}\\

\paragraph*{\textbf{Observability of head-on black-hole mergers.}}
The bottom panel of Fig.~\ref{fig:trajectories} shows the distance at which three of HOC simulations yield $\rho_{\rm opt} \geq 12$ across the Advanced LIGO-Virgo network working at its design sensitivity after averaging over inclination, sky-location, and polarisation angles. These distances are within reach of current BBH observations, so that these sources can in principle be detected.\\ 
\textcolor{black}{While less likely than quasi-circular mergers, HOCs may arise as the result of hierarchical BH mergers \cite{Fragione3,Fragione4} and via gravitational-wave capture \cite{Fragione5}. We also note that near-HOC mergers may occur in resonant interactions between single and BBH, especially between two BBHs, as discussed in~\cite{Ecc099}. Through scattering experiments of BBH interactions~\cite{Ecc099} found that highly eccentric binaries can form within the sensitivity range of Advanced LIGO and Advanced Virgo. The initial eccentricity of those binaries is close to unity (i.e.~nearly head-on) and the mergers typically occur within a second after the system becomes unbound.}\\

\paragraph*{\textbf{Robustness against numerical simulations settings.}}

\begin{figure}
\includegraphics[width=0.49\textwidth]{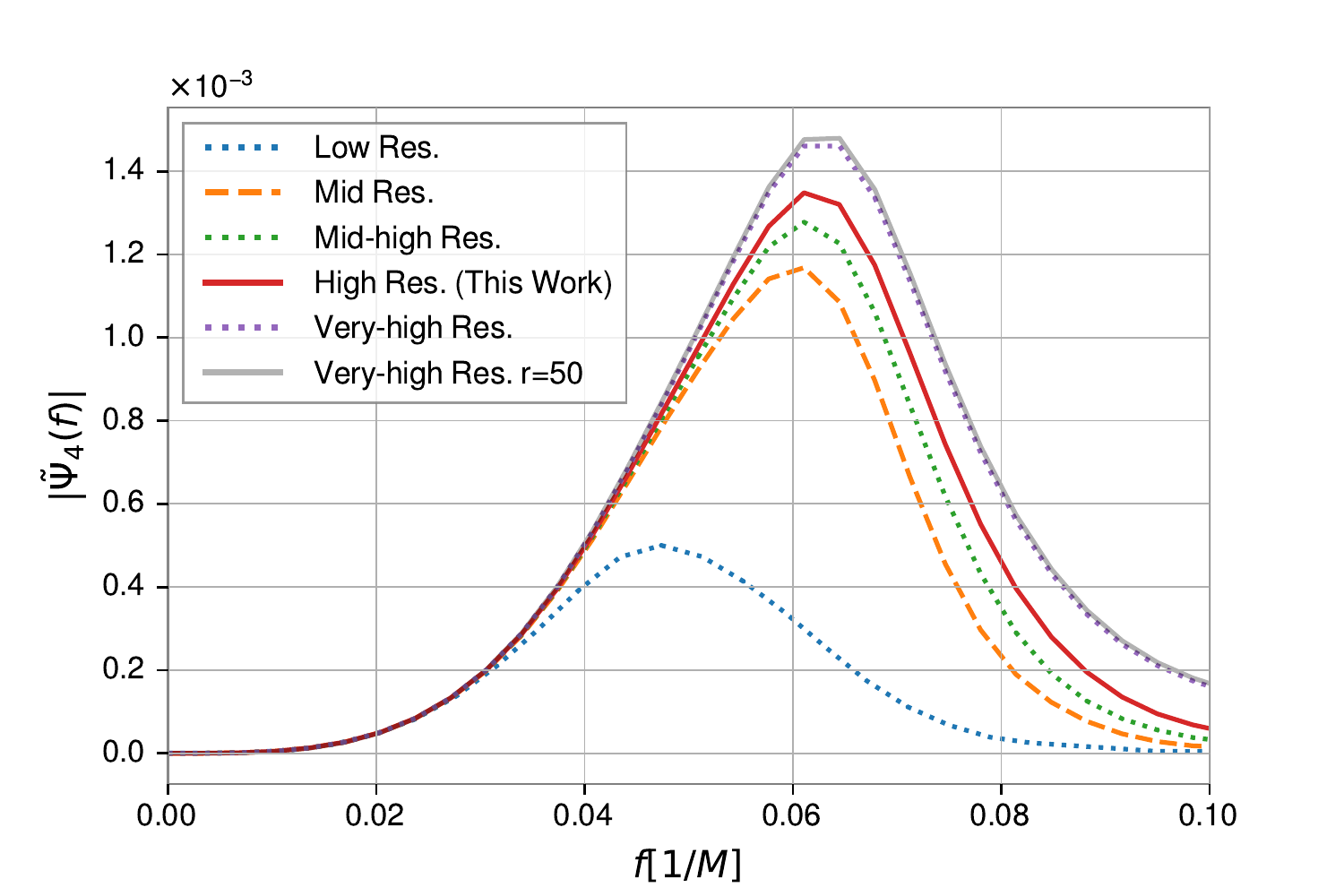}
\caption{The Newmann-Penrose scalar $\Psi_4$, expressed in the Fourier domain, for the three $q=1$ HOCs shown in the bottom panel of Fig.~\ref{fig:converg}, in which waveforms are extracted at a radius of $r=150M$. A clear evolution in the ringdown frequency (at peak) can be noticed as the resolution level increases. The top resolution waveform coincides with that extracted at $r=50M$, for which waveforms for all resolutions are equal (see Fig.\ref{fig:converg}, top panel). Unlike common practice, we show a linear version of this plot instead of a logarithmic one so that the mentioned difference can be easily noticed.
}
\label{fig:ringdown}
\end{figure}

\textcolor{black}{In this section we briefly discuss the accuracy of our numerical simulations and the robustness of our results. Head-on mergers of black holes are among the most studied examples of numerical evolutions of space-times. A convergence analysis of the \texttt{Einstein Toolkit} evolving binary BH mergers can be found in~\cite{Loffler:2011ay}. We show in Fig.~\ref{fig:converg}, the gravitational waveform extracted from a HOC simulation for five different resolutions, extracted at two different radii, namely $r=50M$ (top) and $r=150M$ (bottom). We find between third and fourth order convergence with increasing resolution. Two main aspects can be noticed in this figure. First, it is clear that all resolution levels lead to the same waveform for the case of $r=50M$. However, it can also be noticed that for both extraction radii, a small amount of junk radiation \cite{Higginbotham2019} is present before the true waveform starts, and that this happens closer to the true waveform in the $r=50M$ case than it does for $r=150M$. Therefore, using waveforms extracted at $r=150M$ ensures that the junk radiation happens early enough that we can cut the waveforms in a way that we safely separate the junk radiation from the true emission. The bottom panel shows, however, that in this case the result is significantly more sensitive to the numerical resolution. This is even more visible in Fig.~\ref{fig:ringdown}, where we show the corresponding Fourier transforms for the $r=150$ cases. Here, a clear evolution of the ringdown (peak) frequency can be observed. Note, however, that our highest ``very-high'' resolution waveform is equivalent to the corresponding, well converged, waveform extracted at $r=50M$.}\\

\textcolor{black}{Given the rather strong dependence of the waveforms extracted at $r=150M$ on the resolution of our simulations, it is important to ensure that our results in the main text, which make use the second highest ``high'' resolution, are not affected by this choice. To this end, we repeat our analysis using the ``very-high'' resolution waveform and compare the results with those obtained for the ``high'' resolution case. Table I shows the posterior median values and symmetric $68\%$ credible intervals obtained for our spinning $q=1$ injection, for a total source-frame mass of $M=250M_\odot$ and scaled to an optimal SNR of $\rho_{opt}= 25$, for the ``high'' resolution used in the main text (Fig. 3) and for ``very-high'' resolution. The results obtained are all fully consistent at the $68\%$ level, .i.e., within $1 \sigma$. We therefore conclude that although precision studies assuming very high SNR observations should make use of the highest resolution waveforms, our results are not affected by the numerical resolution of our waveforms.}\\

\begin{table}[t!]
\begin{center}
\begin{tabular}{lcc}
\hline
\hline \\ 
Parameter  & High-res (main text) & Very-high-res \\ \hline
\\[0pt]
Red-shifted total mass & $246^{+15}_{-12}\; M_\odot$ & $237^{+14}_{-11}\, M_\odot$
\\[6pt]
Source-frame total mass & $204^{+11}_{-9}\; M_\odot$ & $199^{+13}_{-9}\,M_\odot$
\\[6pt]
Luminosity distance & $1034^{+127}_{-157}$ Mpc & $955^{+120}_{-150}$ Mpc 
\\[3pt]
Mass ratio & $3.70^{+0.61}_{-0.50}$  & $3.86^{+0.60}_{-0.59}$  
\\[3pt]
Effective spin $\chi_{eff}$ & $-0.46^{+0.10}_{-0.11}$  & $-0.50^{+0.11}_{-0.07}$  
\\[3pt]
Effective precession spin $\chi_{p}$ & $0.37^{+0.13}_{-0.10}$  & $0.30^{+0.14}_{-0.09}$  
\\[3pt]
\hline
\hline
\end{tabular}
\caption{Median values and symmetric $68\%$ credible intervals recovered for our numerically simulated $q=1$ spinning HOC injection in Fig.~3 of the main text, for the case of optimal SNR $\rho_{opt}=25$, at two different resolutions of our numerical simulation. These are labeled as ``high-res'', which is used in the main text, and ``very-high-res'', which we use for cross-check. We obtain results consistent at the $68\%$ level, i.e., at the $\simeq 1\sigma$ level. } 
\label{tab:parameters}
\end{center}
\end{table}

\paragraph*{\textbf{Unequal-mass results.}}

\begin{figure*}
\includegraphics[width=0.51\textwidth]{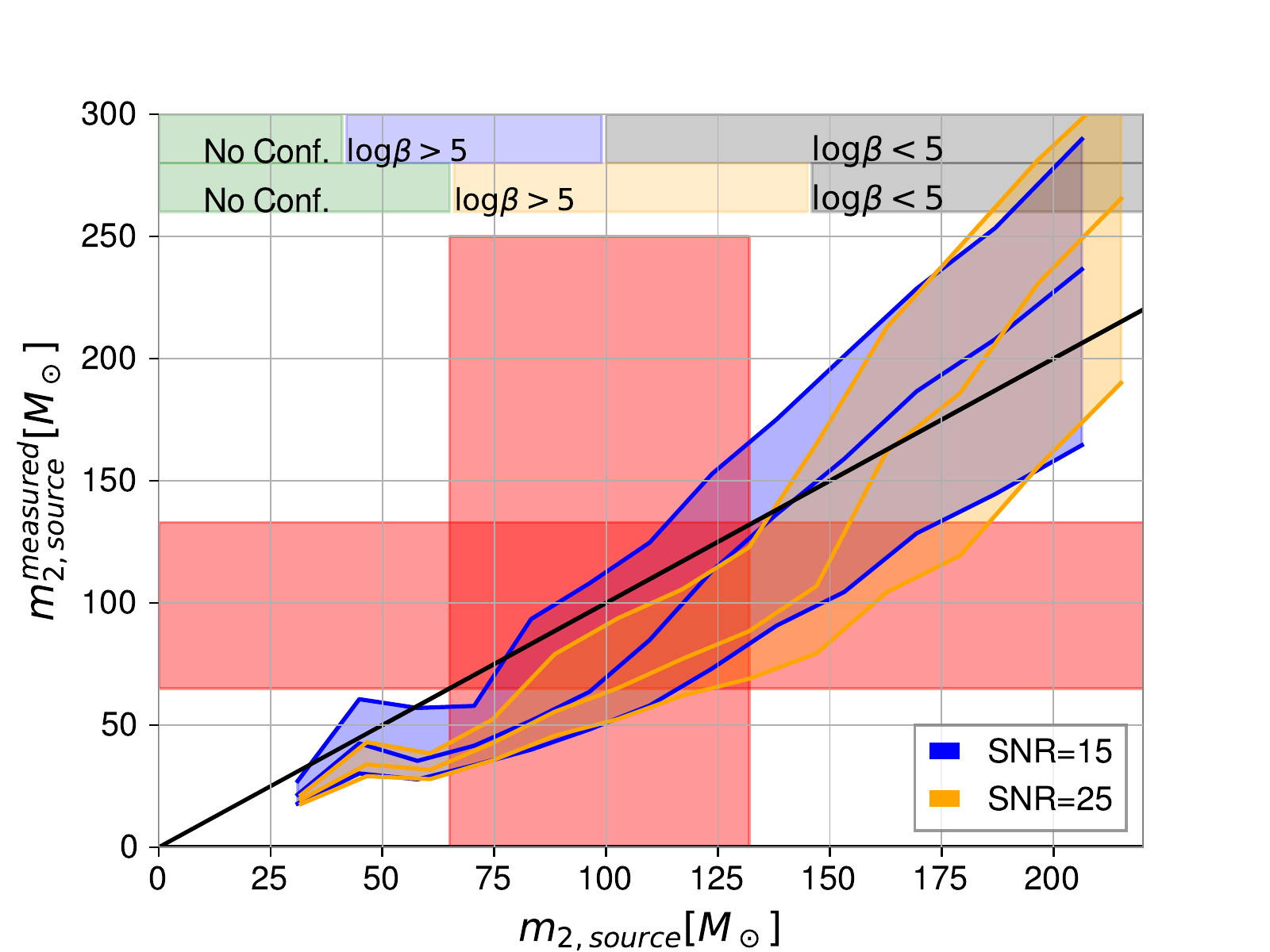}
\includegraphics[width=0.47\textwidth]{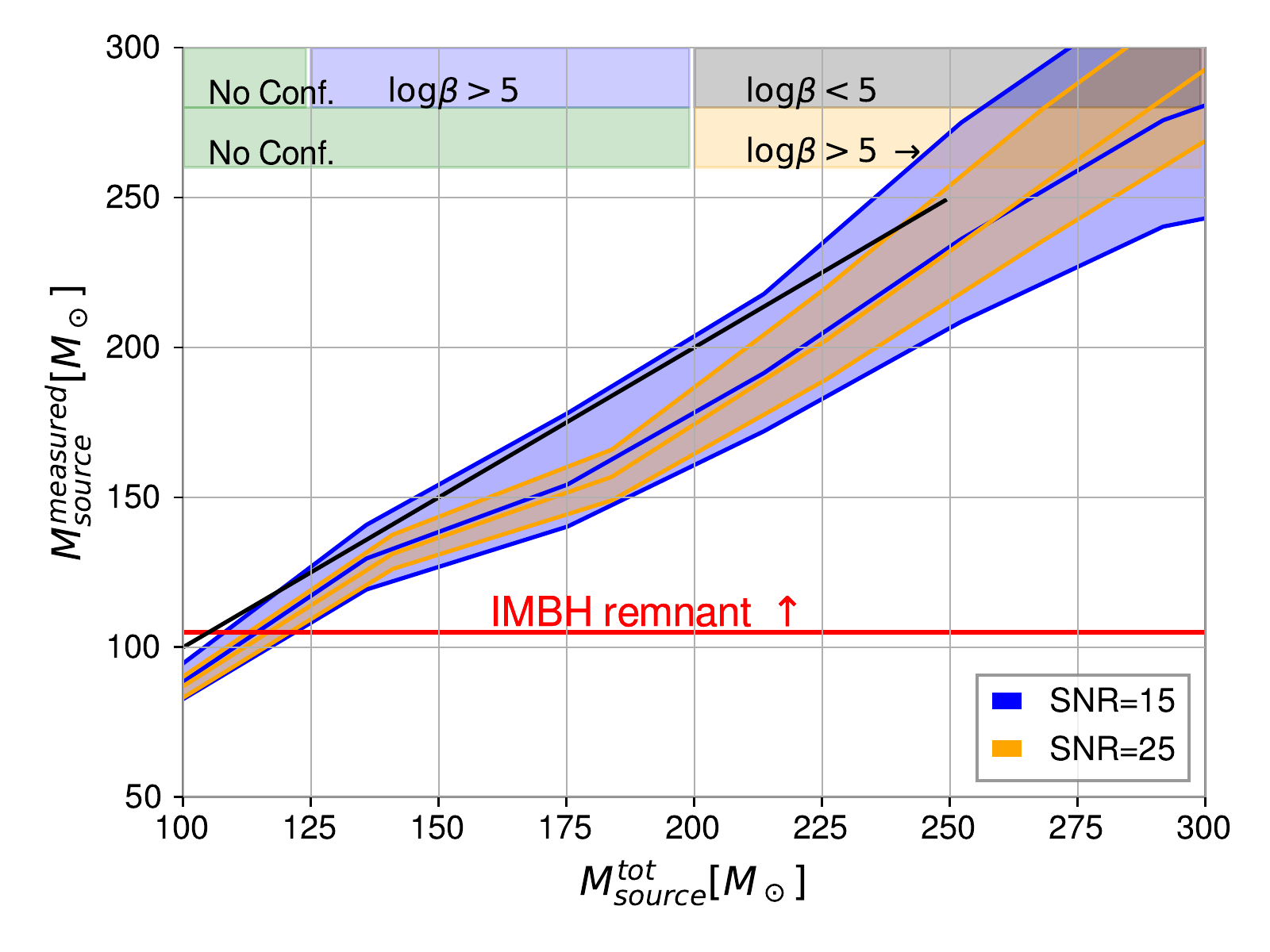}
\caption{
$90\%$ confidence intervals for the secondary mass (left) and total mass (right) of an $q=2$ HOC, measured in the source-frame. The diagonal black line denotes ``zero bias''. The red regions enclose the boundaries of the PISN gap. The top blue and orange regions show the mass ranges in which the source may be confused with a precessing BBH, for $\rho_{\rm opt}=15$ and 25, respectively. Top green regions display ranges in which BBH waveforms have a poor FF, so that the BBH nature of the source can be discarded. Top grey regions denote ranges in which {\tt IMRPhenomD} and {\tt IMRPhenomPv2} fit the head-on injection well enough that no model is preferred. 
}
\label{fig:gap_unequal}
\end{figure*}

\textcolor{black}{In Fig.~\ref{fig:gap_unequal} we reproduce Fig. 5 in the main text for the case of our mass-ratio $q=2$ head-on collision. First, note that the same qualitative phenomenology observed for $q=1$ in Fig.~5 of the main text is present: the left panel shows that for certain range of masses, the lightest black hole of a head-on merger, with mass $m_{2,source}$ inside (outside) the PISN gap, would be interpreted as a black hole with significant support outside (inside) the PISN gap if the signal is analysed using a quasi-circular BBH model. The exact ranges for $m_{2,source}$ in which these situations occur differ from those of Fig. 5 in the main text due to the different mass ratio of the simulation. As in the main text, the right panel shows that for masses below $\simeq 120\,M_\odot$, a remnant IMBH may not be interpreted as such if a quasi-circular BBH origin is assumed, \textcolor{black}{but that comparison with HOC signals would reveal the HOC nauture of the source. Therefore, the IMBH interpretation remains safe.}} 

\bibliography{IMBBH.bib}

\end{document}